\documentclass[rmp,twocolumn]{revtex4}
\usepackage[english]{babel}
\usepackage{amssymb,amsfonts,amsmath}
\usepackage{color}
\usepackage{graphicx}
\usepackage{amsmath}
\usepackage{natbib}
\usepackage{hyperref}
\usepackage{enumerate}

\usepackage{listings}
\newcommand{\EQ}[1]{Eq.~(\ref{eq:#1})}

\newcommand{\FIG}[1]{Fig.~\ref{fig:#1}}

\newcommand{\gt}{g}
\newcommand{\gtfreq}{\gamma}
\newcommand{\afreq}{\nu}
\newcommand{\fcoeff}{\epsilon}
\newcommand{\locus}{s}

\newcommand{\fitprior}{\Phi}

\newcommand{\Author}{Taylor A.~Kessinger${}^{1}$, Alan S.~Perelson${}^{2}$ and Richard~A.~Neher${}^{1}$}
\newcommand{\Affiliation}{${}^{1}$Max Planck Institute for Developmental Biology, 72076 T\"ubingen, Germany, and ${}^{2}$Theoretical Biology and Biophysics, Los Alamos National Laboratory, Los Alamos, NM 87545, USA}

\definecolor{dkgreen}{rgb}{0,0.6,0}
\definecolor{gray}{rgb}{0.5,0.5,0.5}
\definecolor{mauve}{rgb}{0.58,0,0.42}
\begin{document}
\title{Inferring HIV escape rates from multi-locus genotype data}
\author{\Author}
\affiliation{\Affiliation}

\date{\today}

\begin{abstract}
Cytotoxic T-lymphocytes (CTLs) recognize viral protein fragments displayed by major histocompatibility complex (MHC) molecules on the surface of virally infected cells and generate an anti-viral response that can kill the infected cells. Virus variants whose protein fragments are not efficiently presented on infected cells or whose fragments are presented but not recognized by CTLs therefore have a competitive advantage and spread rapidly through the population. We present a method that allows a more robust estimation of these escape rates from serially sampled sequence data. The proposed method accounts for competition between multiple escapes by explicitly modeling the accumulation of escape mutations and the stochastic effects of rare multiple mutants. Applying our method to serially sampled HIV sequence data, we estimate rates of HIV escape that are substantially larger than those previously reported. The method can be extended to complex escapes that require compensatory mutations. We expect our method to be applicable in other contexts such as cancer evolution where time series data is also available.
\end{abstract}

\maketitle

During the first few months of HIV infection, the HIV genome typically
undergoes a series of rapid amino acid substitutions that reduce immune
pressure by cytotoxic T-lymphocytes (CTLs); this process is referred to as CTL escape
\citep{Mcmichael:2009p31614}. The substitutions arise by random mutation and
spread through the viral population by impairing either the presentation of viral epitopes
on the cell surface or the recognition of the viral epitope by T-cell
receptors. Avoiding recognition is an obvious benefit to the mutant virus, but
escape mutations can interfere with processes necessary for virus replication and
infection and thereby reduce the virus' intrinsic fitness \citep{Fernandez:2005p44304,Li:2007p43266,ganusov_fitness_2011,seki_ctl_2012}. The rate at which escape
variants displace the founder sequences depends on both ``avoided killing''
and the fitness cost. To quantify the role of individual CTL clones in
controlling the viral population and the fitness costs associated with escape
mutations, one would like to infer the escape rate associated
with the individual mutations from serially sampled sequence data 
\citep{ganusov_fitness_2011,Asquith:2006p28003}.

\begin{figure}[tbp]
\begin{center}
  \includegraphics[width=0.48\columnwidth]{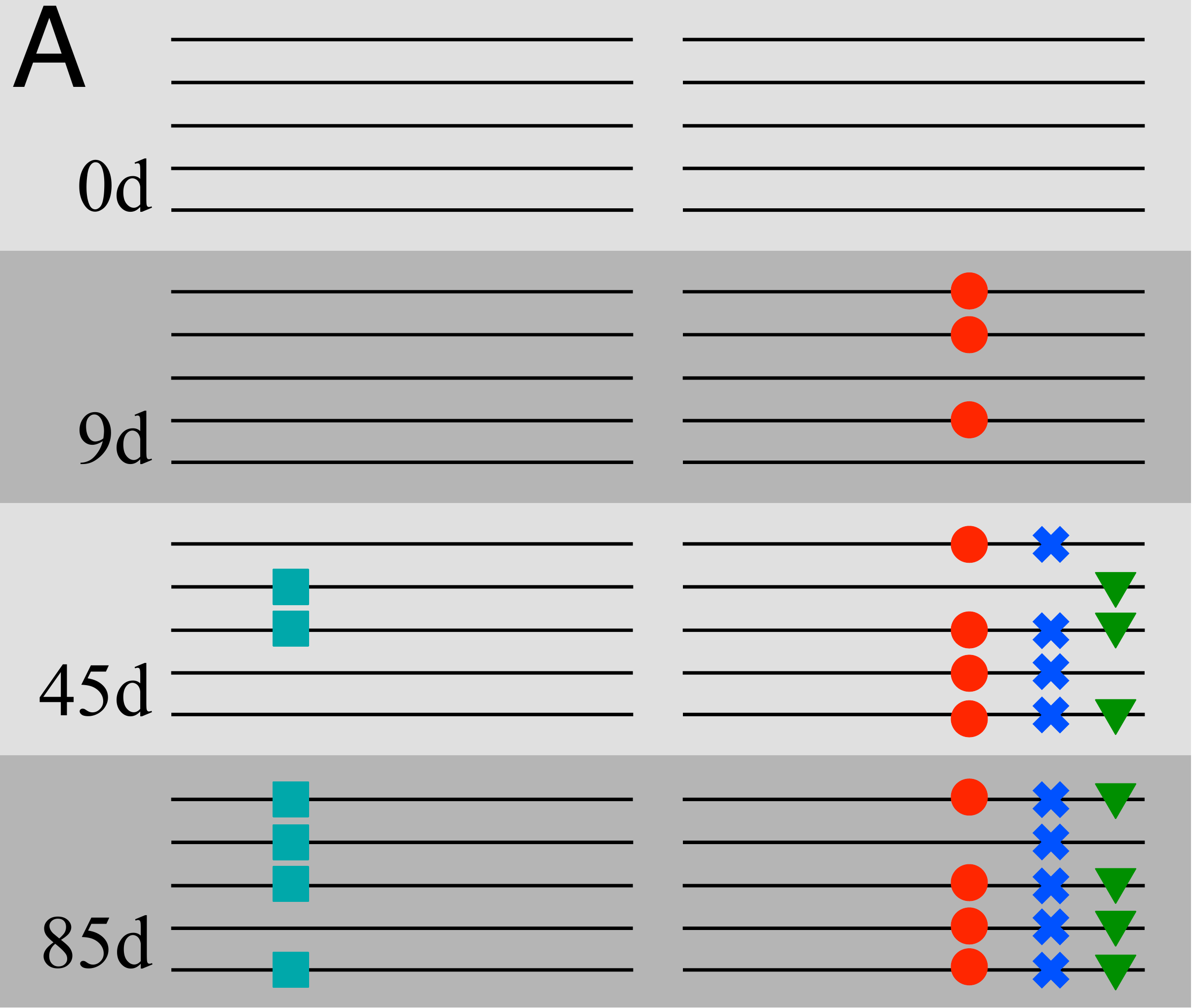}
  \includegraphics[width=0.46\columnwidth]{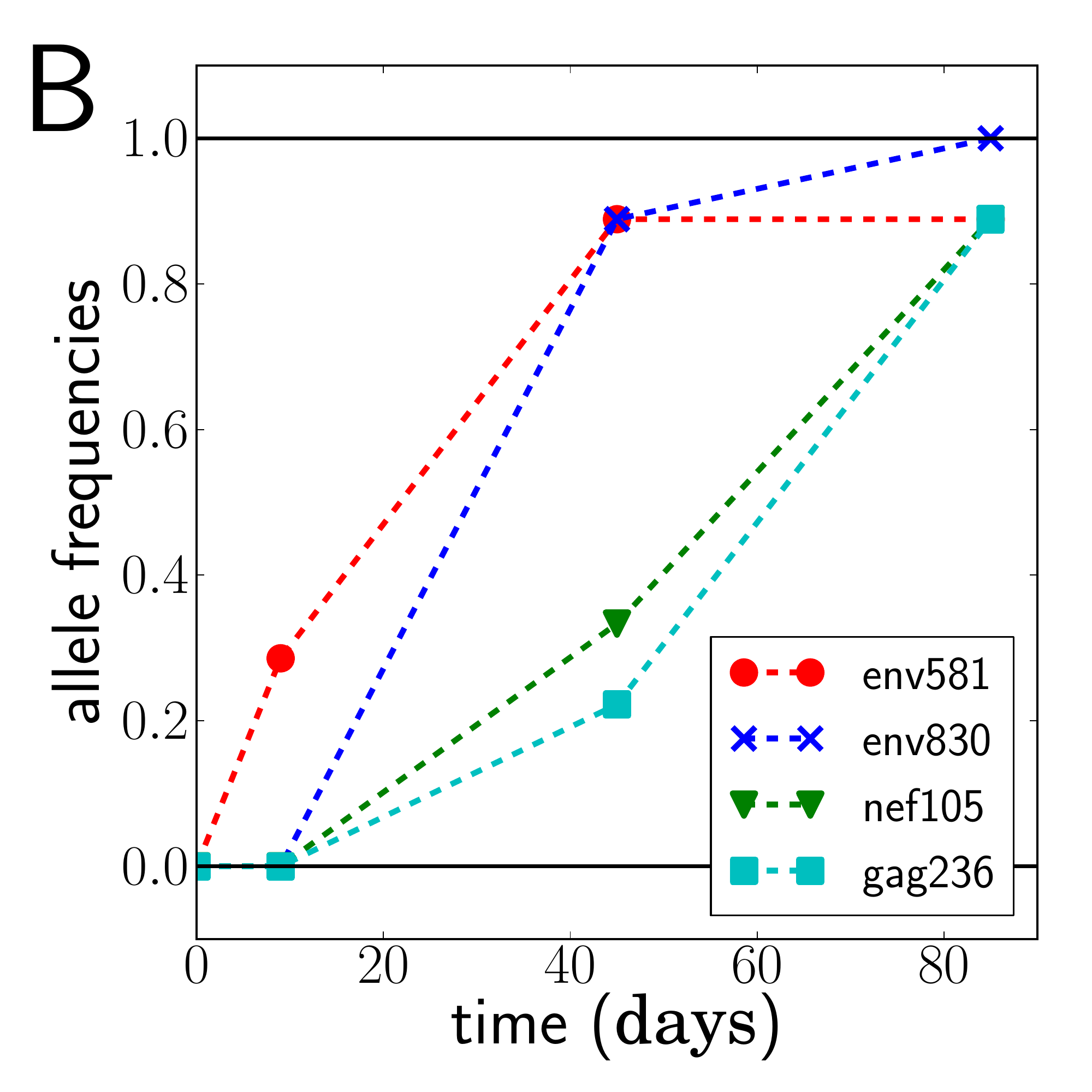}
  \caption[labelInTOC]{Escape from T-cell mediated immunity. The virus
  population in patient CH58 quickly acquires four substitutions. Panel A
shows a sketch of genotypes at the first 4 escape mutations, observed
  at different times; see \citep{SalazarGonzalez:2009p35091,Goonetilleke:2009p42296} 
  for the actual
  data. Panel B shows the frequencies of the mutations in samples of size 7 
  at day 9 and size 9 at days 45
  and 85.}
  \label{fig:data_example}
\end{center}
\end{figure}

With a single escape mutation and dense, deeply sampled data, the escape
rate can simply be estimated by fitting a logistic curve to the time
course of the mutation's frequency
\citep{ganusov_fitness_2011,Asquith:2006p28003}.  The logistic curve has
two parameters: the growth or escape rate and the frequency at the
initial time point. In many cases, however, the data obtained from
infected patients are scarce, and estimating two parameters reliably
from the data is not possible since one needs at least two time points
at which the mutation is at intermediate frequency between 0 and 1
\citep{ganusov_fitness_2011}. Figure \ref{fig:data_example} shows an
example of such time series sequence data from CTL escape during early
HIV infection. Time points are far apart and the sampling depth is
low. Furthermore, it is not the case that only a single escape mutation
is observed; rather, several mutations rapidly emerge in different
places in the viral genome
\citep{Goonetilleke:2009p42296,SalazarGonzalez:2009p35091}. Multiple
escapes imply immune pressure on many epitopes. Since the viral
population and its mutation rate are large
\citep{Perelson:1996p23158,Mansky:1995p38971}, these different escape
mutations will arise almost simultaneously. Initially, these escape
mutations exist in the population as single mutant genomes until they
are combined into multiple mutants by recurrent mutation or
recombination
\citep{ganusov_mathematical_2013,leviyang_computational_2013}. The
competition between viral variants affects the trajectories of
individual escape mutations, so estimating their intrinsic growth rate
by logistic fitting is not accurate. This competition is known as
``clonal interference'' in population genetics. The degree of competition between
genotypes depends on the population size, the mutation rate, and the
recombination rate in HIV populations. The latter-most is rather low
\citep{Neher:2010p32691,Batorsky:2011p40107}, and two strongly selected
mutations in a large populations are more likely combined by additional
de novo mutation than recombination with another rare single mutation.

Here, we develop a strategy for inference that allows one to obtain robust escape 
rate estimates
from the scarce data typical of studies of CTL escape. The inference is based on
explicit modeling of the process of mutation accumulation in the founder
sequence. Thereby, we exploit constraints imposed by the underlying dynamics of
mutation and selection in the high dimensional space of possible genotypes.

Despite the large number of possible genomes that can be formed from
different combinations of escape mutations, we typically observe one or
two dominant genotypes at a time -- at least during the first few month
of the infection. Furthermore, these genotypes dominate only transiently
and are quickly displaced by genotypes with an even greater number of
escape mutations; see \FIG{data_example}. 
These observations agree with results from
ref.~\citep{silva_dynamics_2012}, where a model of acute HIV infection
was used to show that strongly selected escape mutations fix sequentially. 
Note that we don't assume a
particular sequence of dominant genotypes a priori. Instead, we observe a
sequence of dominant genotypes and try to infer the evolutionary
scenario that most likely gave rise to this sequence of genotypes. While
we model only these genotypes, many minor variants certainly exist. But
only those dominant variants that are likely to give rise to the future
populations need to be modeled accurately. Later in infection, the viral
population is very diverse and cannot be analyzed using our method.

Given a data set from early infection, it is typically straightforward to
define a series of dominant genotypes that likely have arisen through
step-wise accumulation of mutations. Note that most likely all escape mutations constantly
arise in different combinations, but typically only one combination
rises quickly enough to dominate the population. This dominant genotype
is then in most cases the source for the next dominant genotype. Later
in infection, however, recombination is sufficiently frequent that no
dominant genotype exists and mutations can spread simultaneously.

In Ganusov et al \cite{ganusov_mathematical_2013}, a framework for
multi-locus modeling of CTL escape is presented. Building on this
framework, we explicitly model the transition from one dominant genotype
to another, which is a good approximation of the dynamics for rapid CTL
escape in acute infection. The restriction to dominant genotypes
captures the interference between escapes at different epitopes while
avoiding the need to solve the full multi-locus problem.

We will first define a model of the dynamics of escape mutations. This model
serves a two-fold purpose: it defines the parameters we would like to
estimate from the data and provides us with a computational tool to
investigate how the accuracy of the inference depends on sampling depth and
frequency, as well as how sensitively it depends on the values of parameters
such as mutation rates or the population size. We reanalyze existing CTL
escape data and find that accounting for multi-locus effects in a finite population
results in higher estimates of the escape rates.

\section{Results}
\subsection{Model} In the majority of sexually transmitted HIV infections, a single 
``transmitted/founder" virus initiates the new infection
resulting in an initially homogeneous viral population 
\citep{SalazarGonzalez:2009p35091,keele_identification_2008}.
However, as HIV replicates in its new host, mutations
accumulate. Mutations within or in proximity to CTL epitopes can reduce 
immune pressure by facilitating the avoidance of CTL recognition. 
While one often observes several escape mutations within a single
epitope \citep{Fischer:2010p40314,henn_whole_2012}, we do not differentiate between different mutations
within the same epitope and model $L$ epitopes that can be either be
mutant or wild-type.
Assuming that the escape at multiple epitopes has additive effects,
$\fcoeff_j$, the growth rate (birth rate minus death rate) of a genotype is given by
\begin{equation}
\label{eq:growth_rate}
F(\gt,t) = F_0(t) + \sum_i \fcoeff_i \locus_i
\end{equation}
where $\gt = \{\locus_1, \ldots, \locus_L\}$ specifies
the genotype. Here, $\locus_i=0$ corresponds to a wild-type epitope at locus $i$, 
whereas $\locus_i=1$ signifies escape at that epitope. $F_0(t)$ accounts
for a genotype independent modulation of 
the growth rate. The latter could, for example, be due to variable numbers of
target cells \citep{petravic_cd4+_2008,ganusov_estimating_2006}. $F_0(t)$ controls the total population size, while the differences
between genotypes are accounted for by $\sum_i \fcoeff_i \locus_i$ and result in
differential amplification of some genotypes over others. The
$\fcoeff_i$ are the escape rates that we would like to estimate from the data and should
be interpreted as the net effect of avoided killing and the possible fitness costs associated
with the mutation; see \textit{e.g.}~Ganusov et al
\cite{ganusov_mathematical_2013}. The fitness costs are modulated 
by the overall growth rate of the viral population and could therefore
be slightly time dependent. We neglect this complication.

Within our model, mutations arise at a rate
$\mu$ per base per generation. This rate can be epitope dependent. 
Motivated by the frequent template switching of HIV reverse transcriptase 
\citep{Levy:2004p23309}, our general
model of the HIV population includes recombination, which is assumed to occur
with rate $r$. In the event of recombination, all $L$ epitopes are
reassorted, but an explicit genetic map could be implemented as well.

We implemented our model as a computer simulation in Python using the population genetic
library FFPopSim \citep{zanini_ffpopsim:_2012}. 
The simulation stores the population $n(\gt,t)$ of each of the $2^L$
possible genotypes. In each generation, the expected changes of the $n(\gt,t)$
due to mutation, selection, and recombination are calculated. The population of
the next generation is then sampled from the expected genotype frequencies
$\gtfreq(\gt,t) = n(\gt,t)/N$.
The size of the population, $N$, can be set at will each generation. In this way, up to
15 epitopes can be simulated for 1000 generations within seconds to minutes. 

\begin{figure*}[tpb]
\begin{center}
  \includegraphics[width=0.62\columnwidth]{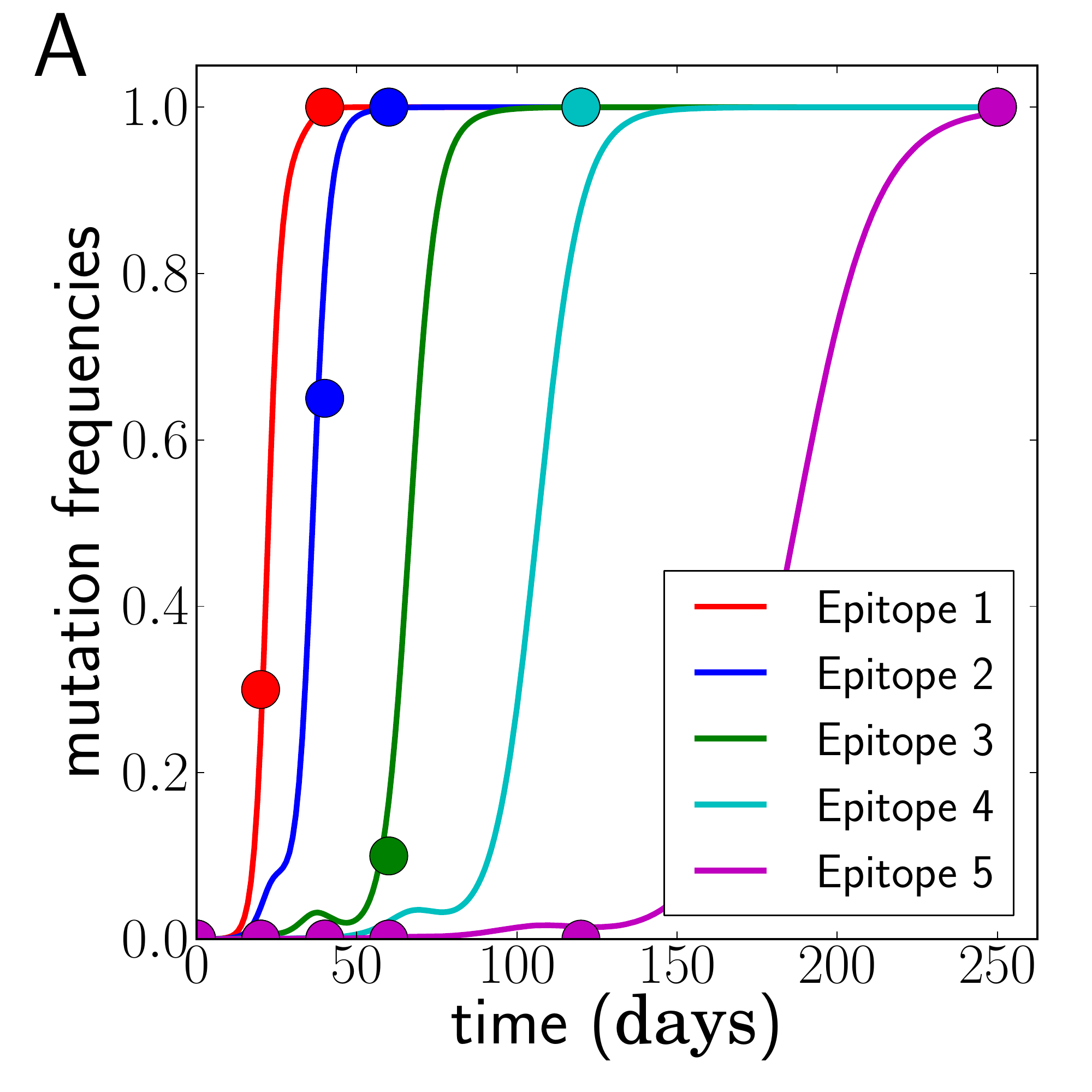}
  \includegraphics[width=0.62\columnwidth]{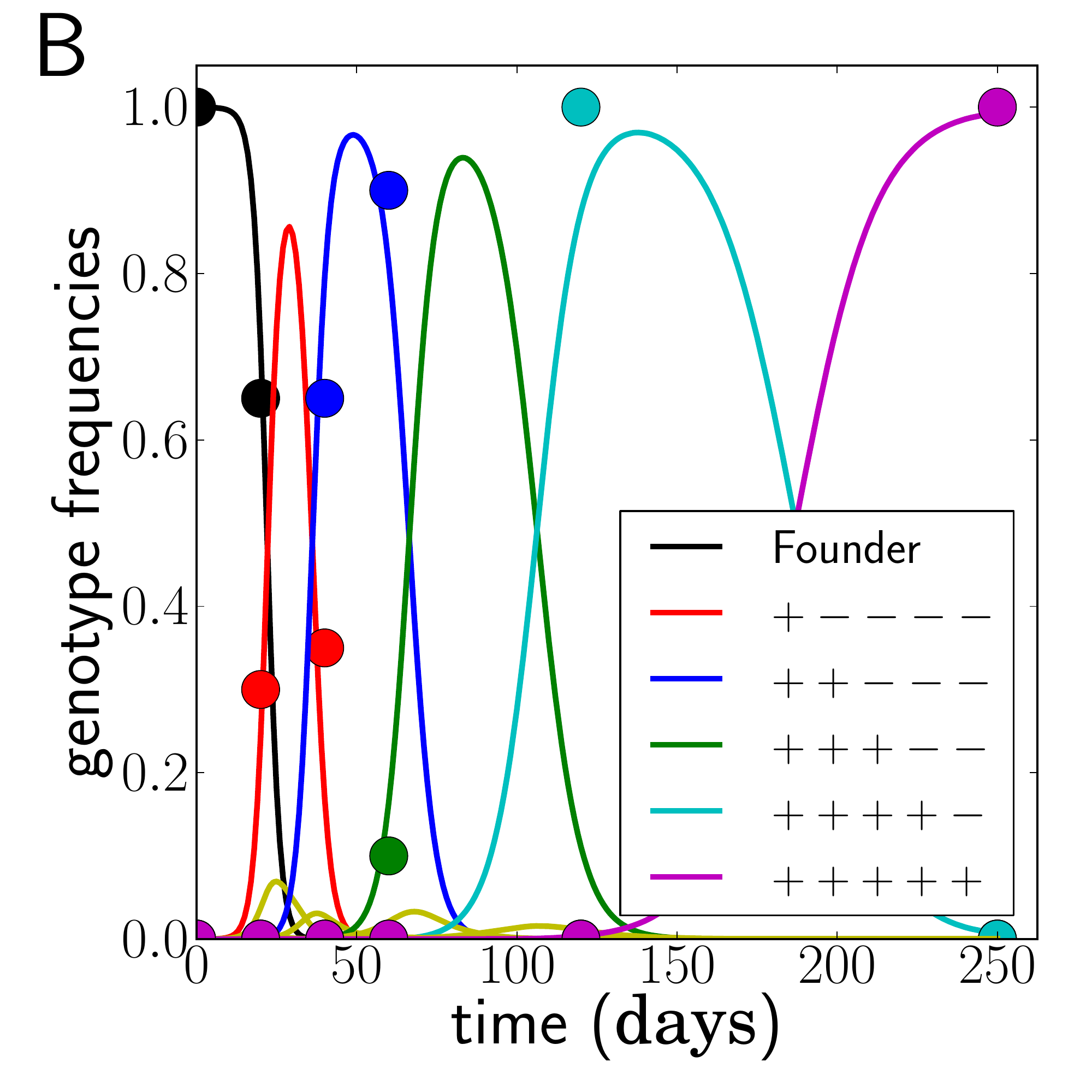}
  \includegraphics[width=0.62\columnwidth]{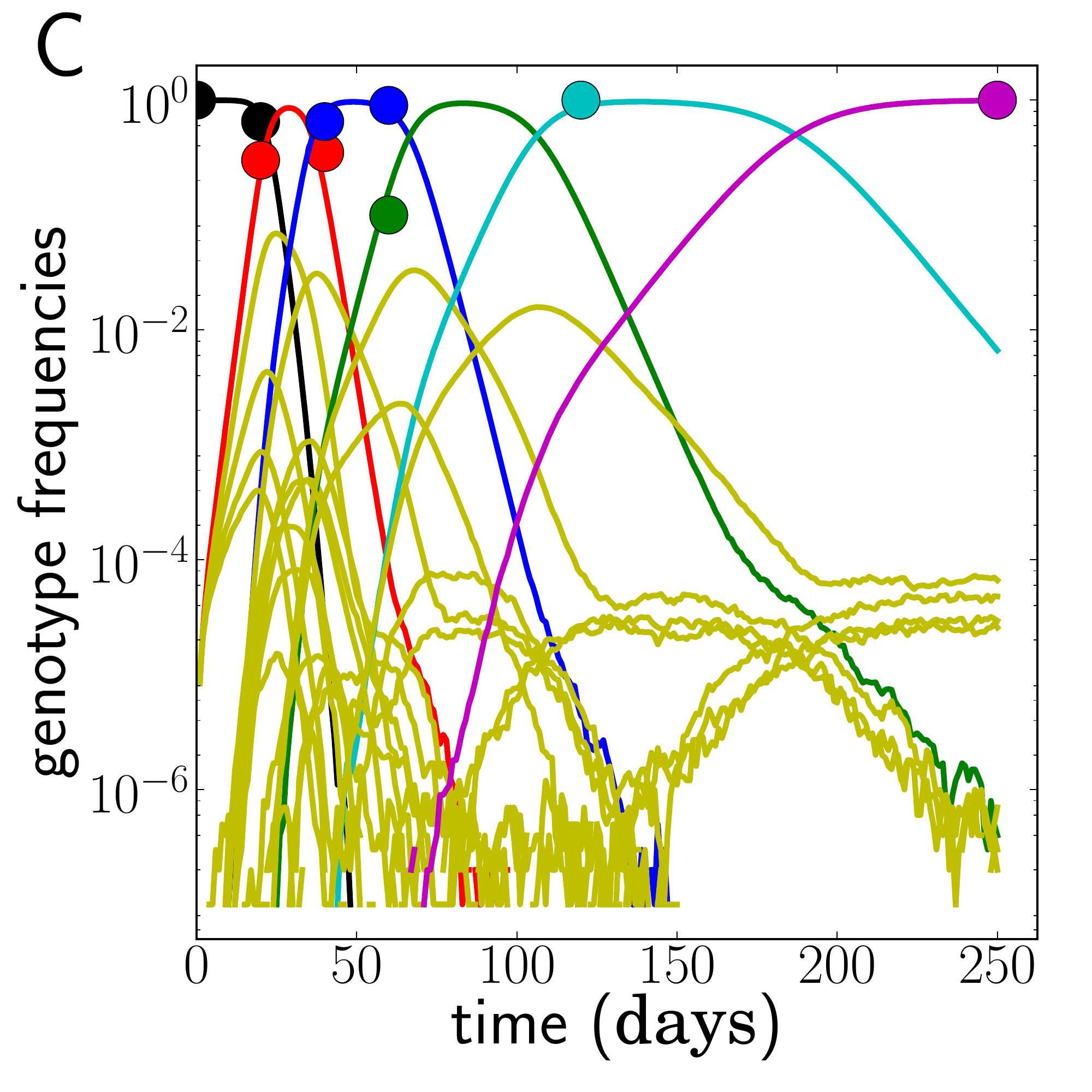}
  \caption[labelInTOC]{Example of simulated escape mutations spreading 
    through the population. (A) Even though all epitopes are targeted from $t=0$, 
    escape mutations spread sequentially. The mutation frequency in a
    sample of size 20 at different time points is indicated by colored
    dots. (B) The rising mutation
    frequencies are associated with the rise and fall of multilocus
    genotypes. The founder virus is first replaced by a dominant single
    mutant, which itself is replaced by a double mutant and so forth. 
   Note, however, that the virus population explores many combinations
   of mutations but that these minor variants never reach appreciable frequency.
  This is best seen in panel (C), where all 32 genotype
  frequencies are shown on a logarithmic scale. These rare variants are rarely
  sampled, and their noisy dynamics suggests that little information can be
  gained from them. Here,
$N = 10^7$, $\mu = 10^{-5}$, and $r = 0$ and escape rates are
    $\fcoeff_j = 0.5, 0.4, 0.25, 0.15, 0.08$ per day.}
  \label{fig:example}
\end{center}
\end{figure*}

A typical realization of the population dynamics is shown in
\FIG{example}, where we have assumed a generation time of one day. 
As expected, the population is dominated by one genotype at a time.
Furthermore, the mutations accumulate in decreasing order of escape rate, and the new
dominant genotype arises from the previous by incorporation of the mutation with
the largest escape rate available. There are, however, many minority genotypes which
are rarely observed. Figure \ref{fig:example}C shows the frequencies on a
logarithmic scale, where the minor variants are visible. We use these
simulations to test the accuracy and robustness 
of the inference procedure developed below.

Of the many possible genotypes that are present at any moment, only a
small fraction is likely to be observed in a small sample and to be
relevant in the future. Simulations and data suggest that the dominant
genotypes accumulate mutations one by one -- this greatly simplifies the
task of estimating escape rates from the data. Instead of considering
the dynamics of all possible genotypes ($2^L$), we will restrict the
inference to a chain of genotypes, each containing one additional
mutation compared to its predecessor.

The best estimates for the HIV generation time are $d=2$ days
\cite{markowitz_novel_2003}, while estimates of escape rates are
typically given in units of inverse days rather than generations. For
simplicity, we simulate our model assuming one generation per day and
state all rates in units of 1/day. Our results are insensitive to the
choice of the generation time. Doubling the generation time has similar
effects to dividing the population size by 2, as this keeps the strength
of genetic drift constant.

\subsection{Inferring the escape rates}
Suppose we have obtained sequence samples of size $n_i$ at different time points
$t_i$ and each of these samples consists of different genotypes $\gt$ present in
$k(\gt,t_i)$ copies. If the actual frequencies of the those genotypes at
different times are $\gtfreq(\gt,t_i)$, the probability of obtaining the sample at
$t_i$ is  given by the multinomial distribution
\begin{equation}
P(\mathrm{sample}) = \frac{n_i!}{\prod_\gt
k(\gt,t_i)!}\prod_{\gt}\gtfreq(\gt,t_i)^{k(\gt,t_i)}
\end{equation}
If the underlying dynamics were deterministic, the frequencies $\gtfreq(\gt,t)$
would be unique functions of the model parameters we want to estimate. In that
case we could use Bayes' theorem, choose suitable priors, and determine the
posterior distribution of the parameter values. However, both the
model and the actual viral dynamics are stochastic, and ``replaying''
the history would result in different trajectories. Furthermore, most of the $2^L$ possible
genotypes remain unobserved. This leaves us with the choice of either
some type of approximate Bayesian computation that compares repeated simulations
of the model with appropriate summary statistics \citep{sunnaker_approximate_2013} 
or a reduced description
of only the observed genotypes, with the stochasticity captured by nuisance parameters
\citep{basu_elimination_1977}.

We opt for the latter and model only those genotypes that dominate the
population.  We label these genotypes by the number of escape mutations
they carry, e.g., $\gt_1$ carries the first escape mutations, $\gt_2$
the first and the second, and so forth.  The 
frequency of a genotype is
affected by stochastic forces only while it is very rare. If the
genotype is favored, it will rapidly rise to high frequency, and the
stochastic effects will no longer be relevant. It is therefore
convenient to summarize the stochastic behavior by the time, $\tau$,
at which its frequency crosses the threshold to essentially
deterministic dynamics.  Since the dynamics is deterministic after this
``seed time'', all the (unobserved) stochasticity can be accounted for by
an appropriate choice of the seed time
\citep{Kepler:1995p26819,Desai:2007p954}. For each of the dominant
escape variants, $\gt_j$, with $j=1$ to $j=L$ escaped epitopes, we define a
seed time $\tau_j$ to accommodate the stochastic aspects of the escape dynamics. 

\begin{figure}[tp]
\begin{center}
  \includegraphics[width=0.8\columnwidth]{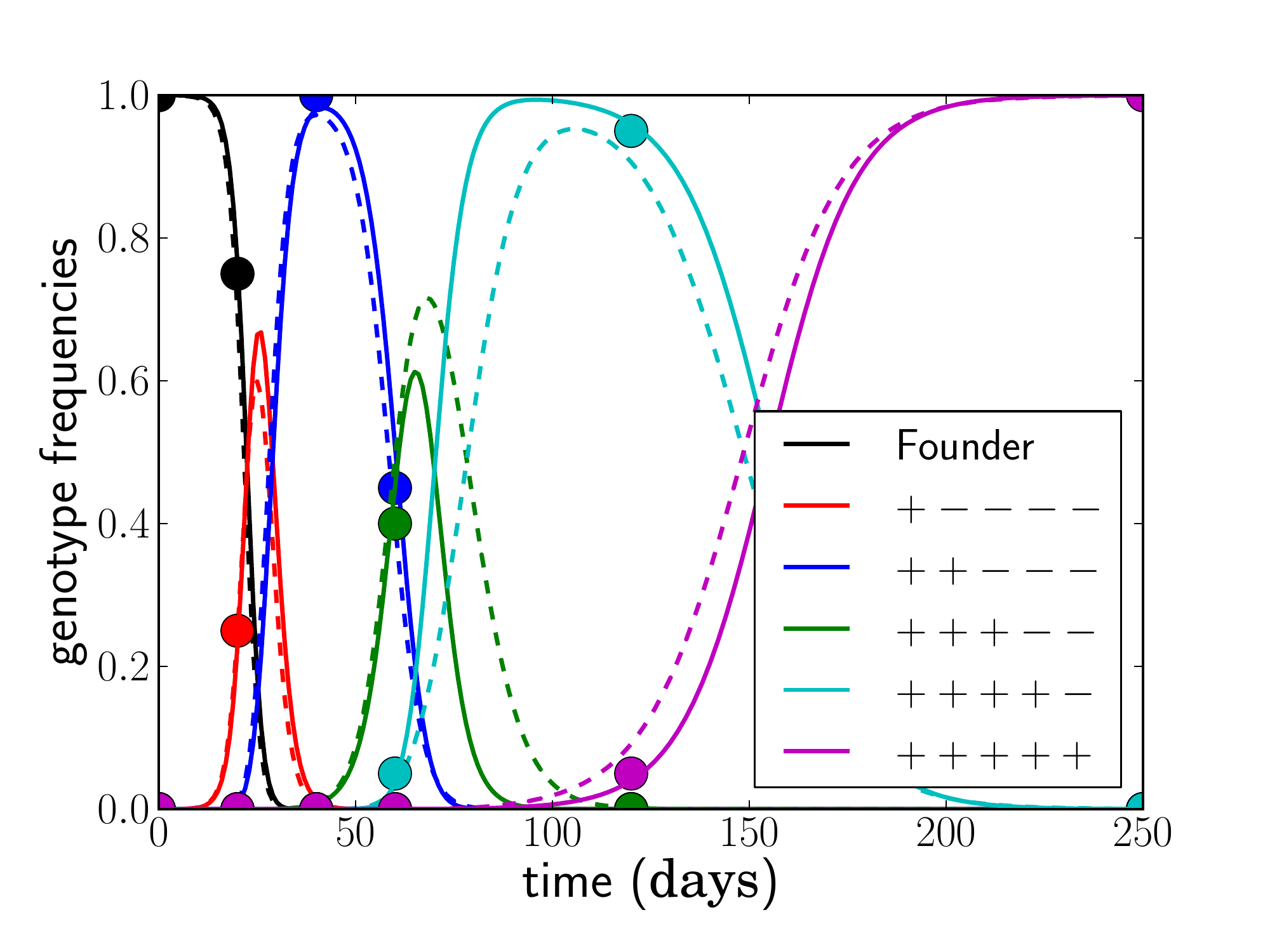}
  \caption[labelInTOC]{The deterministic model parameterized by seed times
  $\tau_j$ for the $L$ dominant genotypes and the escape rates of
  epitopes $\fcoeff_i$ 
  (solid lines) captures the dynamics of the stochastic model accurately
  (dashed lines). The trajectories (and seed times) vary from run to run. In this run,
$N = 10^7$, $\mu = 10^{-5}$, and $r = 0$ and the escape rates are
    $\fcoeff_j = 0.5, 0.4, 0.25, 0.15, 0.08$ per day.}
  \label{fig:model_vs_full}
\end{center}
\end{figure}

After crossing the deterministic threshold, the population frequencies 
of the dominant genotypes evolve according to
\begin{equation}
\label{eq:gt_dynamics}
\dot \gtfreq_j(t) = F(\gt_j,t)\gtfreq_j(t) +\mu [\gtfreq_{j-1}(t) - \gtfreq_{j}(t)]
\end{equation}
if $t>\tau_j$. Conversely, $\gtfreq_j(t)=0$ for $t<\tau_j$. The growth
rate $F(\gt_j,t)$ of genotype $j$ is the sum of the escape
rates $\fcoeff_k$ of the epitopes $k=1,\ldots,j$ and the density
regulating part $F_0(t)$; compare to~\EQ{growth_rate}. The escape rates are what we would like to
estimate.  The seed time, $\tau_j$, corresponds to the time
at which a genotype with all escape mutations up to mutation $j$ first
establishes\footnote{There is a brief period after the initial
  production of the mutation during which the dynamics is stochastic and
  the initial mutant establishes only with a probability roughly equal
  to $\epsilon d$, where $d=2$ days is the generation time
  \citep{markowitz_novel_2003}. However, we find $\epsilon d\approx 1$ and 
  ignore this complication.}.  At the seed time, we initialize the
genotype frequency at $\gtfreq_j(\tau_j)=N^{-1}$. If seed times are
chosen appropriately, this model provides a very accurate description of
the frequency dynamics of the dominant genotypes in the full stochastic
model; see \FIG{model_vs_full}.

At face value, the deterministic model has two parameters per epitope --
one escape rate and one seed time. The
seed times, however, are quite strongly constrained by basic facts of the
evolutionary dynamics. The genotype $\gt_j$ carrying mutations
$i=1,\ldots,j$ arises with
rate $\mu N(t)\gtfreq_{j-1}(t)$ from the genotype $\gt_{j-1}$ carrying only $j-1$ mutations.
This means it is unlikely that genotype $j$ arises early while $\gtfreq_{j-1}(t)$ is still
very small. However, once the previous genotype $j-1$ is common, genotype $j$ 
is produced frequently. The distribution of the time at which the first copy 
of genotype $j$ arises is given by the product of the rate of production and the 
probability that it has not yet been produced. The latter is the negative exponential 
of the integral of the production rate up to this point. Hence, the distribution of the 
seed time $\tau_j$, given the trajectory of the previous genotype $\gtfreq_{j-1}$,
is given by
\begin{equation}
\label{eq:seedtimes}
Q(\tau_j | \gtfreq_{j-1}(t)) \approx \mu N(\tau_j)\gtfreq_{j-1}(\tau_j) e^{-\mu
\int_0^{\tau_j} N(t)\gtfreq_{j-1}(t)\,dt} \ .
\end{equation}
Since the $\gtfreq_j(t)$ are uniquely specified by $\{\tau_k,\fcoeff_k\}_{k=1,\ldots,L}$, we can
write the posterior probability of the parameters as 
\begin{equation}
\label{eq:LH}
P(\{\fcoeff_j,\tau_j\})  \propto \prod_{i} P(\mathrm{sample}_i|\Theta)\prod_j
Q(\tau_j|\Theta)U(\fcoeff_j),
\end{equation}
where $\Theta = \{ \fcoeff_k,\tau_k\}_{k=1\ldots L}$ and $U(\fcoeff_j)$
is our prior on the escape rates. 
We employ a Laplace prior $U(\fcoeff) = \exp(-\fitprior\fcoeff)$  parameterized by $\fitprior$ 
favoring small escape rates. The prior regularizes the search for the minimum and results in
conservative estimates of escape rates.

\subsection{Obtaining maximum likelihood estimates} 
Finding the set of escape rates and seed times that maximizes the posterior probability
can be difficult due to multiple maxima and ridges in the high
dimensional search space, and uncertainty remains.
To ensure that the global optimum will be reliably discovered, we exploit the
sequential nature of the dynamics and use the fact that earlier escapes
strongly affect the timing of the later ones, but not vice versa. Thus adding
genotypes with an increasing number of mutations one at a time 
results in a reasonable initial guess on top of which a global true multi-locus search can be
performed.

\begin{figure}[htp]
\begin{center}
  \includegraphics[width=0.8\columnwidth]{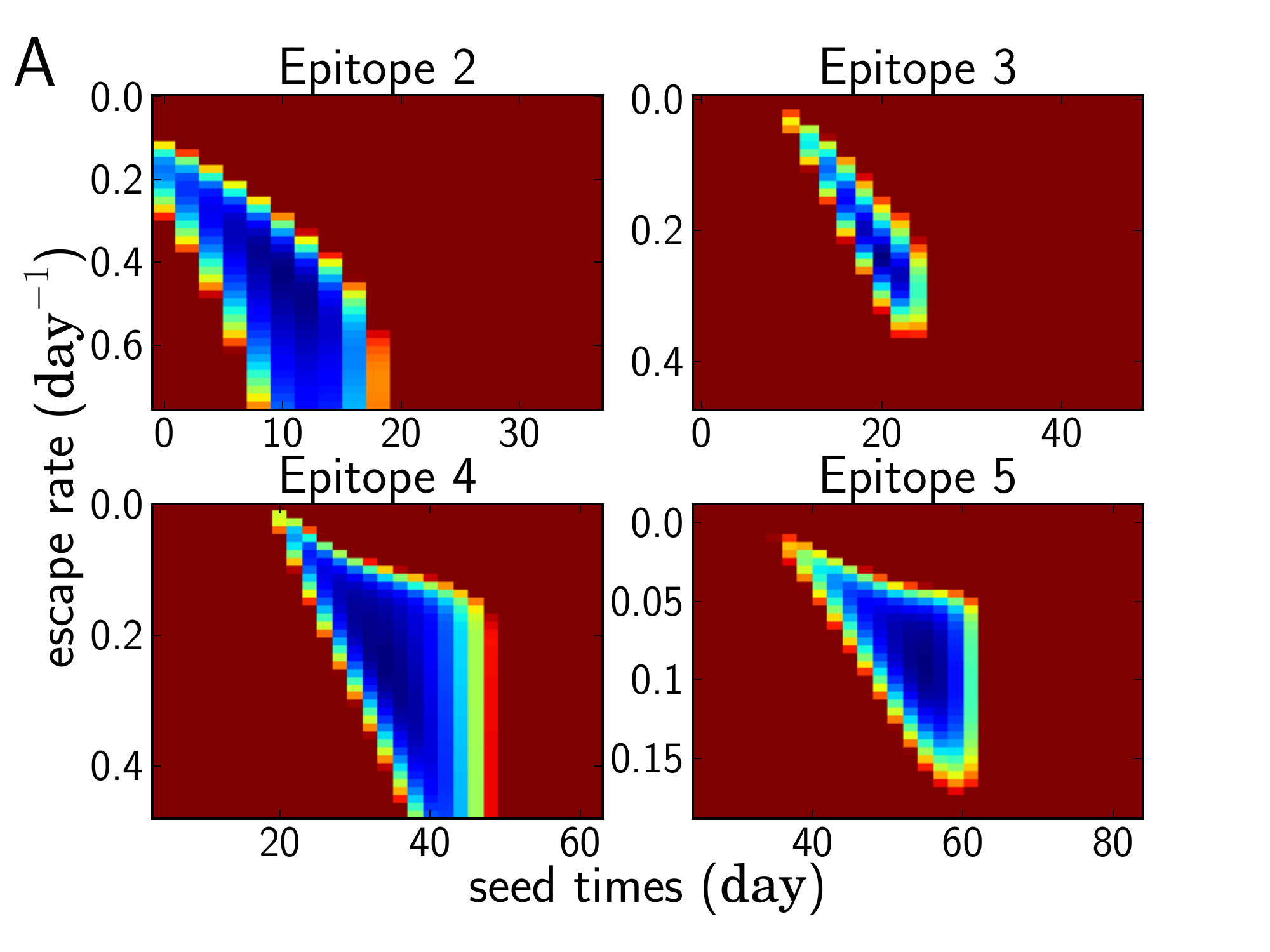}
  \includegraphics[width=0.8\columnwidth]{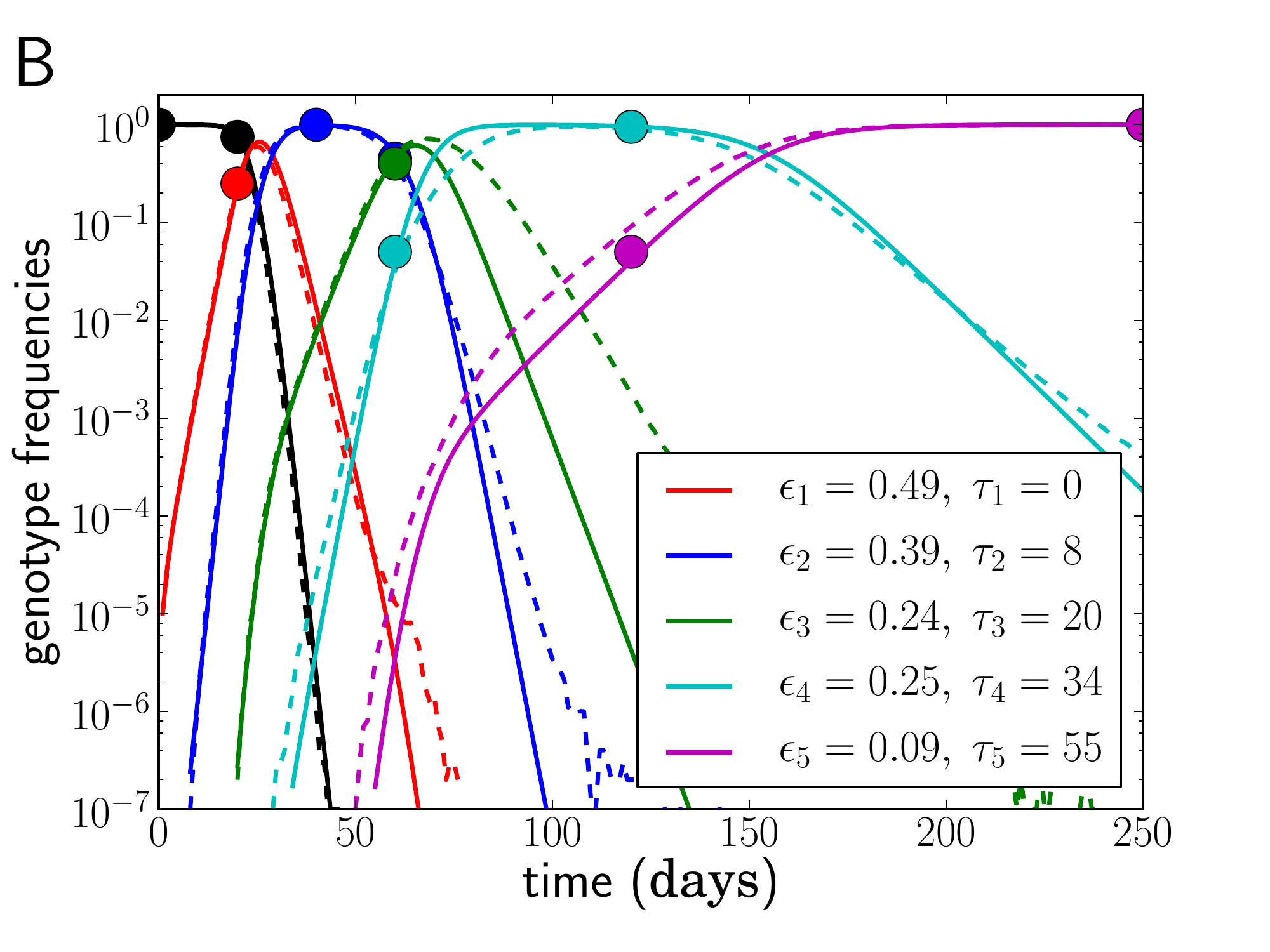}
  \caption[labelInTOC]{Adding epitopes one by one is a feasible and
    reliable fitting strategy. Assuming we know the population was
    homogeneous at $t=0$, there is only one free parameter for the first
    epitope, which is easily determined.  For all subsequent epitopes,
    we need to determine the seed time $\tau_j$ and the escape rate
    $\fcoeff_j$. In panel A, the negative log posterior probability of
    these parameters is shown for each of the epitopes. The surface
    typically exhibits a single minimum.  Panel B shows the genotype
    frequencies of the founder virus and the dominant escape variants 
    (solid lines: model fit, dashed lines: actual
    simulated trajectories). The estimated escape rates of individual
    epitopes and the seed times of genotypes containing all escape mutations up to
    $j$ are given in the legend. Only the samples indicated by balls (20
    sequences at each time point) were used for the estimation. In this
    run, $N = 10^7$, $\mu = 10^{-5}$, $r = 0$, and the escape rates
    $\fcoeff_j = 0.5, 0.4, 0.25, 0.15, 0.08$ per day.}
  \label{fig:sequential_fitting}
\end{center}
\end{figure}

We have implemented such a search in Python, while the computationally expensive
calculation of the posterior probability is implemented in C. The code infers
parameters as follows:
\begin{itemize}
  \item Fit the first escape assuming $\tau_1=0$ by a
  simple one dimensional minimization. This assumes that single mutants
  are already present in the population, consistent with the large viral population
  size present by the time a patient has been identified as HIV-1 infected \citep{coffin_hiv_1995,perelson_dynamics_1997}.
  \item Add additional epitopes successively by mapping the entire two-dimensional
  posterior distribution $P(\fcoeff_j,\tau_j)$ at fixed $\{\fcoeff_k,\tau_k\}$
  for $k<j$. This step is illustrated in \FIG{sequential_fitting}A.
  \item Refine the estimates through local optimization via gradient descent, Monte
  Carlo methods, or local exhaustive search. The resulting parameters and
  trajectories are shown for one example in \FIG{sequential_fitting}B.
  \item Generate posterior distributions by Markov chain Monte Carlo (MCMC). 
\end{itemize}

This procedure is described in more detail in the methods section. Fitting a five epitopes
takes on the order of a minute on one 2011 desktop machine (Apple iMac i7 2.93 GHz). Generating
the local posterior distribution by MCMC takes roughly 20 minutes for $10^{6}$ steps.

\subsection{Comparison to simulated data}
To evaluate the accuracy and reliability of our inference scheme, we
performed true multi-locus stochastic simulations using FFPopSim (see
methods) and sampled genotypes from the simulation at a small number of
time points. Time points and sample sizes were chosen to mimic patient
data. We then inferred parameters from this ``toy" data and compared the
result to the actual values. When interpreting these comparisons, it is
important to distinguish two sources of error.  First, limited sample
size and sampling frequency will incur errors due to inaccurate
estimates of the actual genotype frequencies from the sample. The second
source of uncertainty is an inappropriate choice of model or model
parameters. Such inappropriate model choices might include wrong
estimates of the population size or mutation rates, the presence or
absence of recombination, or time variable CTL activity.

We generate data assuming escape rates $\fcoeff_j = 0.5, 0.4,  0.25,
0.15, 0.08$ per day and sample the population on days $t_i=0,20,40,60,120,250$. An
example of such samples is shown in \FIG{example}. Note that each genotype
is typically only sampled at a single data point; it easily happens that
a genotype is hardly seen at all. We therefore expect all inferences to be quite
noisy as is the case with patient data.

\subsubsection{Sample size and sampling frequency dependence} 
With more frequent and deeper sampling, inferring the model parameters is
expected to become simpler. Indeed, as soon as each genotype is sampled more than
once at intermediate frequency, one can estimate its growth advantage simply
from its rate of increase. This is the rationale behind previous studies such as
\citep{ganusov_fitness_2011,Asquith:2006p28003}. In many data sets, however, this
condition is not met. By constraining the seed time based on the evolutionary
trajectory of the previous escape, our method is able to produce a more accurate
reconstruction of parameters with less data.

\begin{figure}[htp]
\begin{center}
  \includegraphics[width=0.98\columnwidth,type=pdf,ext=.pdf,read=.pdf]
{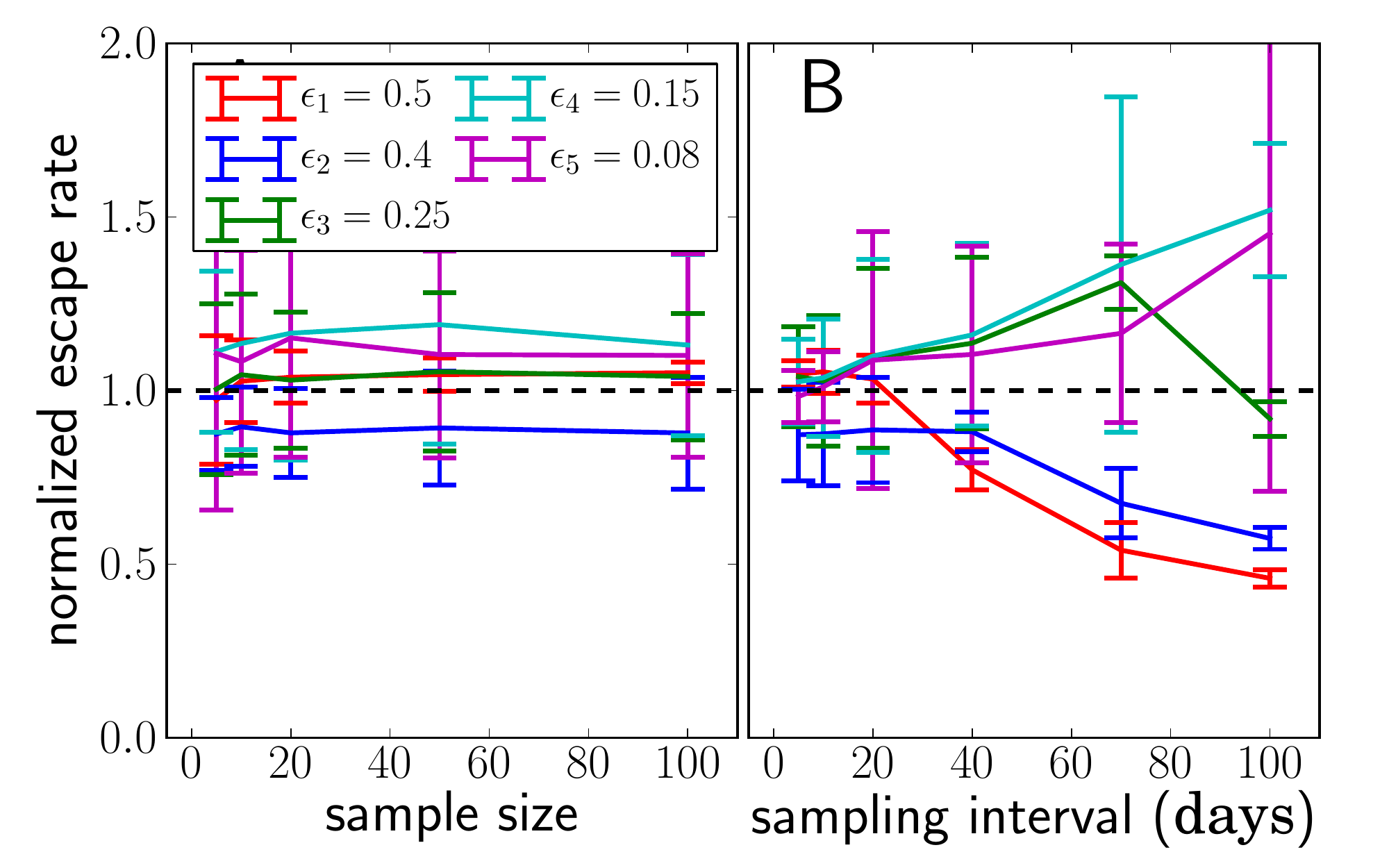}
  \caption[labelInTOC]{The dependence of the accuracy of inference on sample sizes (panel A)
  and sampling intervals (panel B). The actual normalized 
  escape rate is 1.0 and is shown by the dashed line. Sample size only moderately 
  affects the
  accuracy, while sparse sampling (every 40 days in this example) leads to
  serious loss of accuracy. Sample size is
  $n=20$ when sample intervals are varied, and sampling times are as
  illustrated in \FIG{example} when sample size is varied. The plots show the
  mean $\pm$ one standard deviation. The actual values of the escape rates simulated
are shown in the legend (same on both panels). In each run,
$N = 10^7$, $\mu = 10^{-5}$, and $r=0$. Mean and standard deviation at
each point are calculated from 100 independent simulations. }
  \label{fig:sample}
\end{center}
\end{figure}

Figure \ref{fig:sample} shows the estimates obtained as a function of the sampling
frequency and sample size. Increasing the sample size improves the estimates
only moderately, while increasing the sampling frequency leads to substantial
improvements.

\subsubsection{Model deviations}
\begin{figure*}[htp]
\begin{center}
  \includegraphics[width=1.99\columnwidth]{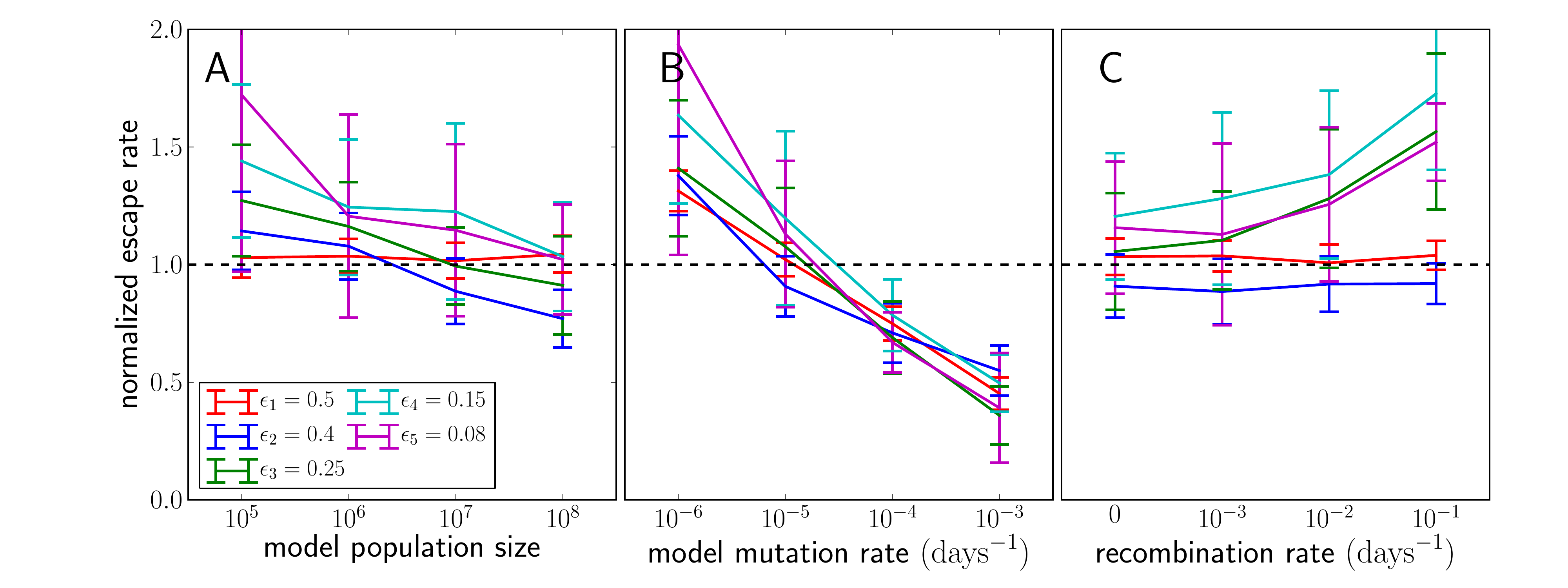}
  \includegraphics[width=1.99\columnwidth]{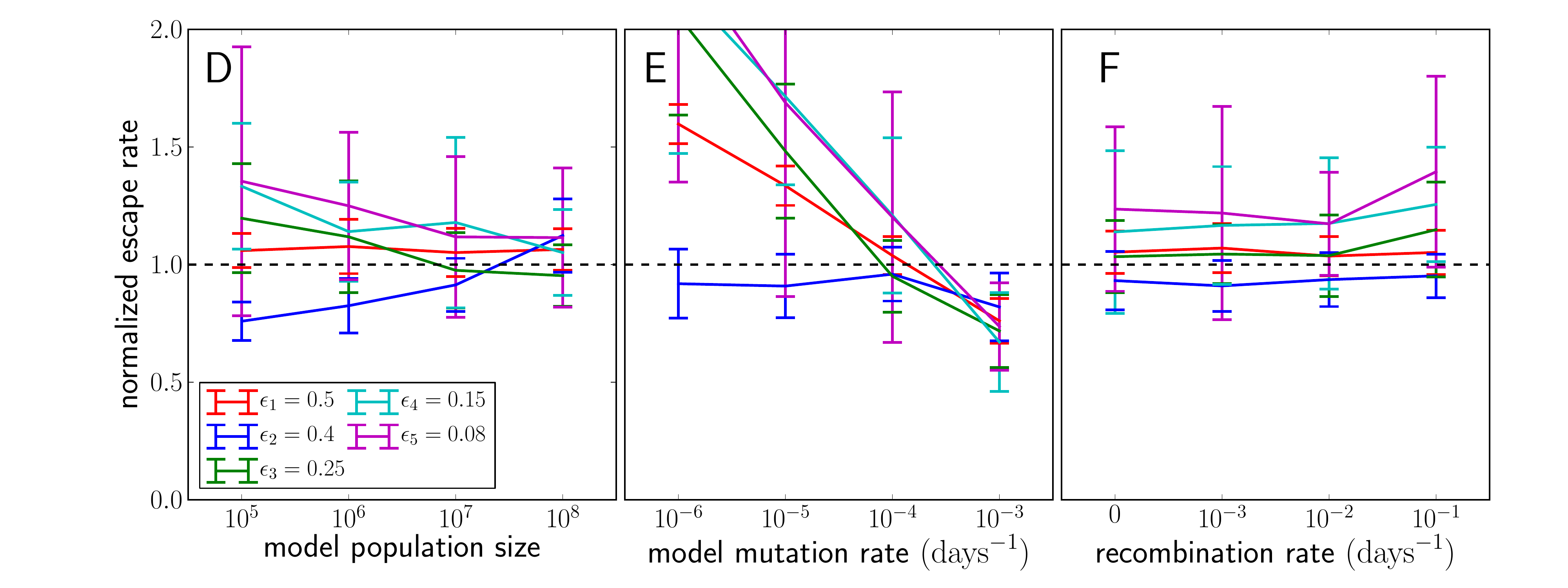}  \caption[labelInTOC]{The effect of assuming the wrong population parameters on the escape rate estimates. To quantify the robustness against wrong assumptions, we simulate escape dynamics with parameters different from those assumed in the escape rate estimation. Panels A-C show simulations with $N=10^{7}$ and $\mu=10^{-5}$ per day, while panels D-F use a ten-fold higher mutation rate $\mu=10^{-4}$. In panels C and F, the simulated recombination rate varies as shown.
  A \& D) Assuming a too small population size
  results in estimates that are too large. The effect is more pronounced at lower mutations rates. 
B) Similarly, if the mutation rate is assumed too large, the estimated seeding of multiple mutants occurs too early and the estimates of escape rates are too low. Note that assuming the correct rates ($\mu=10^{-5}$ in B and $\mu=10^{-4}$ in F) results in unbiased estimates. 
 C \& F) If the population recombines, the actual seed times are smaller
  than those estimated by the fitting routine. To compensate for the
  shorter time interval during which the escape variant rises, the
  estimates of escape rates are larger than the actual escape rates at least at 
low mutations rates. For high mutations rates, recombination is less important since additional mutations are more efficient at producing multiple mutants than recombination. Mean and standard deviation at
each point are calculated from 100 independent simulations.  }
  \label{fig:modeldeviation}
\end{center}
\end{figure*}

The population size and the mutation rate explicitly enter our model through the
seed time prior, but we rarely know these numbers accurately. Hence we
need to understand how inaccurate assumptions affect our estimates.
If we assume that $N\mu$ is larger than it really is, our inference
method will favor seeding subsequent
genotypes too early, which in turn results in erroneously small estimates of escape rates. We varied
$N$ and $\mu$ and observed the expected effect on the estimates as shown in
\FIG{modeldeviation}. The dependence on $\mu$ is stronger than that on $N$, since
the effect of a larger population size is partly canceled by the longer time
necessary to amplify the novel mutation to macroscopic numbers.
However, even the dependence on $\mu$ is rather weak, and
changing $\mu$ ten-fold only changes estimates of escape rates by $\pm 50\%$.
The underlying reason is that the seed times depend primarily on the logarithm
of $N\mu$. $Q(\tau_j|\gtfreq_{j-1}(t))$ (see \EQ{seedtimes}) peaks when $N\mu
\gtfreq_{j-1}(t)\approx 1$. Since $\gtfreq_{j-1}(t)$ is growing
exponentially, the position of the peak changes only logarithmically
with the prefactor $N\mu$. Changes in $\mu$ also affect the dynamics
through the initial rise in frequency of novel genotypes due to recurrent
mutations; see \EQ{gt_dynamics}.

Another factor that affects seed times is recombination. HIV recombines
via template switching following the coinfection of one target cell by
several virus particles \citep{Levy:2004p23309}. In chronic infection,
coinfection occurs with a frequency of about 1\%
\citep{Neher:2010p32691,Batorsky:2011p40107}. Recombination is not
modeled in the seed time prior of our inference method but can speed up
escape by combining escape mutations at different epitopes. As a result,
if recombination is present, seeding tends to happen earlier than our
prior would suggest. If the model assumes that seeding occurs later than in reality,
there is less time for an escape variant to grow to its observed
frequency. Hence the estimated escape rate (growth rate) is larger than
the actual escape rate to compensate for the shorter time.  In
\FIG{modeldeviation}, we compare the estimates obtained by applying our
inference method to simulation data with recombination. Recombination
starts to have substantial effects once coinfection exceeds a few
percent. Recombination primarily affects the incorporation of more
weakly selected mutations and can be ignored for very strongly selected
CTL escape mutations. Recombination also has negligable effects if the
mutation rates is large as is seen in panel F of \FIG{modeldeviation}.

\subsubsection{Unobserved intermediates and compensatory mutations}
\begin{figure}[htp]
\begin{center}
  \includegraphics[width=0.48\columnwidth]{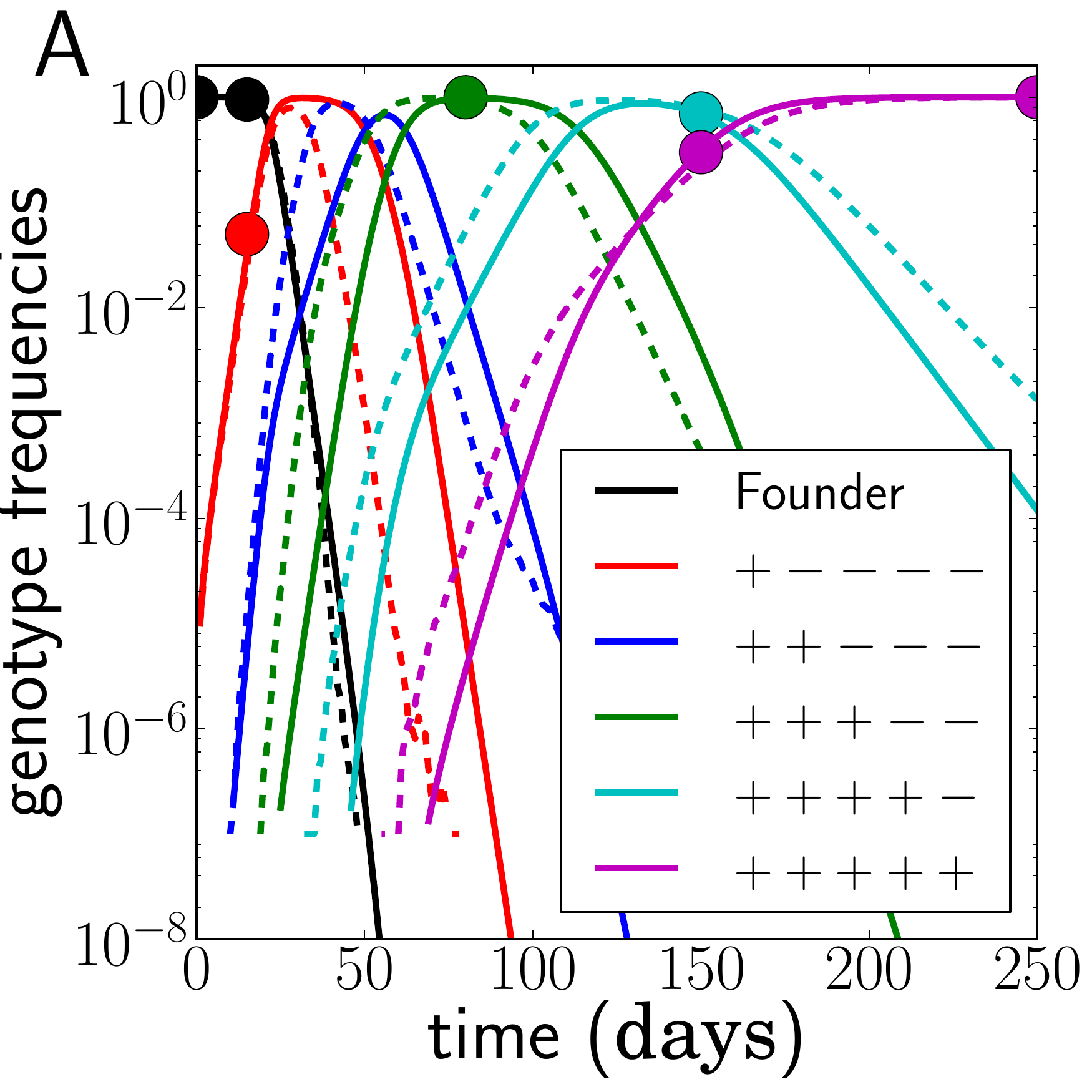}
  \includegraphics[width=0.48\columnwidth]{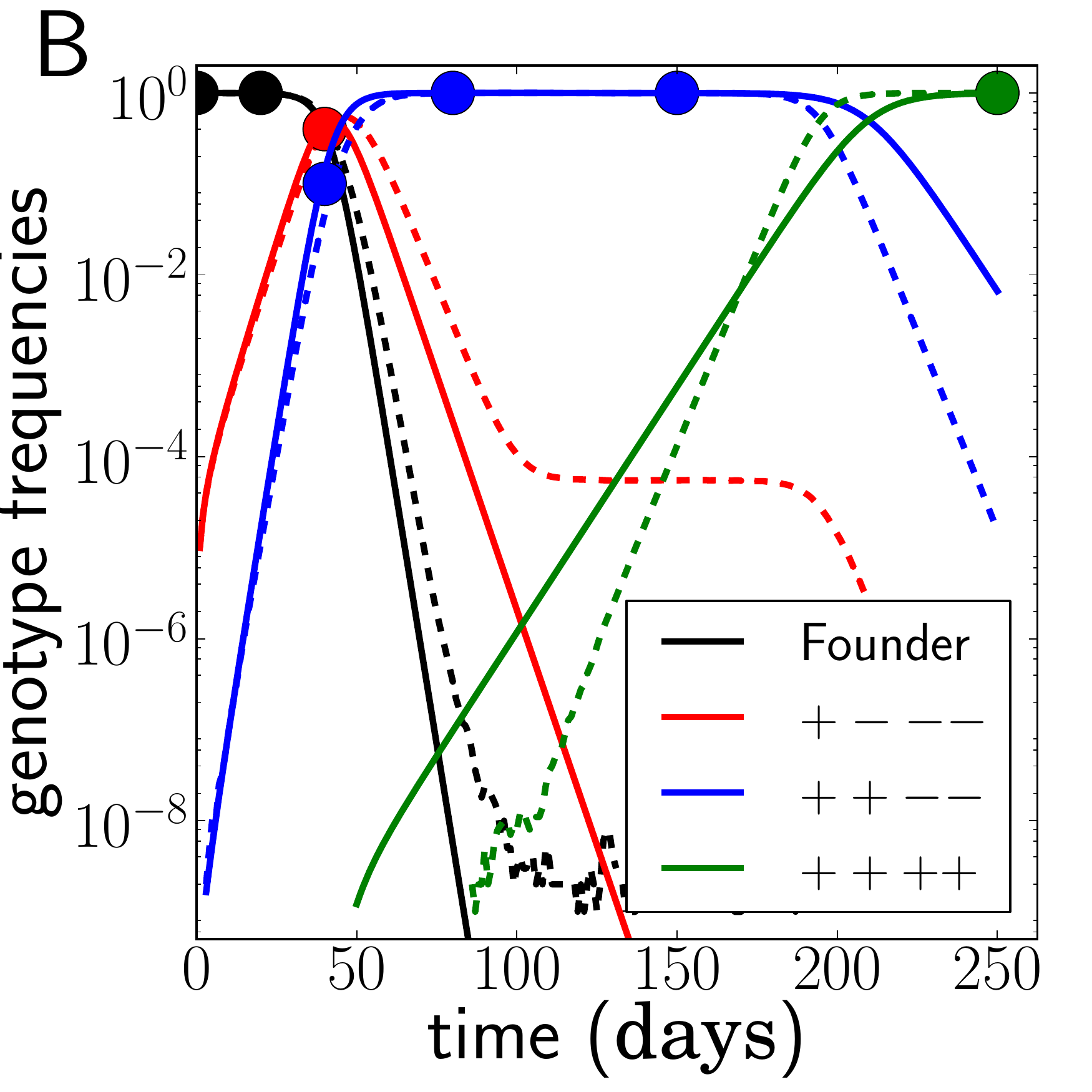}
  \caption[labelInTOC]{Unobserved intermediates and compensatory mutations.
  Panel A shows a scenario where the genotype with only 2 escape
  mutations (blue)
  was not observed even though this genotype was transiently at high frequencies.
  We fit this scenario by assuming both mutations have the same escape rate but
  occur sequentially ($N=10^{7},\, r=0,\, \mu=10^{-5}$). 
  Panel B shows a scenario where escape mutations 3 and 4
  only occur together and any genotype containing only one of the two mutations
  is not viable. Hence the effective mutation rate into the genotype is
  $\mu^2=10^{-10}$ and the waiting time for this genotype is longer. Note that
  the population size is $N=10^9$ in this example ($r=0,\, \mu=10^{-5}$). The last escape only appears
  once the previous escape mutations have reached frequency one, and the seeding
  time is quite variable.}
  \label{fig:degenerate}
\end{center}
\end{figure}

The time intervals between successive samples are sometimes too large to observe
the accumulation of single mutations, so the dominant genotype at one
time point differs by more than one mutation from the previous. This can arise
for two reasons. First, one or several unobserved genotypes may have transiently been at
high frequency but been out-competed by later genotypes before the next sample
was taken.
Second, one escape might have required more than one mutation, for 
example because
single mutants are not viable and a compensatory mutation is needed
\citep{read_stochastic_2012}. Both scenarios can be accounted for in our scheme
and are illustrated in \FIG{degenerate}.

Unobserved, but individually beneficial, intermediate genotypes can be included
by assuming they all have the same escape rate and were seeded one from the
other. There is not sufficient information to estimate more than an average
escape rate for all of them. For a given set of sampled frequencies,
the estimated escape rates increase as more and more intermediates
are assumed. Such unobserved intermediates are common in the data from infected
individuals analyzed below.

Compensatory mutations and ``multiple-hit'' escapes can be accounted for by
replacing the single site mutation rate in \EQ{seedtimes} by the effective rate at which the viable escape mutant
appears. In the simplest case where all intermediate
states are lethal and mutations are independent, this rate is simply the probability $\mu^k$,
where $k$ is the number of mutations needed. In other cases, the rates to
multiple hits can be calculated using branching process approximations
\citep{weissman_rate_2009,Neher:2011p42539}. The choice of the relevant
effective mutation rate for complex escapes must be made on a case-by-case
basis. The effective mutation rate of a multiple-hit escape will often be low
enough that its seed time is not very well constrained. If, for example, the
population size is $N=10^8$ and the effective mutation rate is $10^{-10}$, the
seed time distribution has a width of more than 100 days. Given this weak
constraint, more data are required in order to estimate the escape rate accurately; see
\FIG{degenerate}.

\subsection{Immune escape in HIV-infected patients} 
CTL escape was characterized in detail in the
patients CH58, CH40, and CH77
\citep{Goonetilleke:2009p42296, SalazarGonzalez:2009p35091} and further analyzed
in Ganusov et al \cite{ganusov_fitness_2011}. Sequences were obtained by single
genome amplification followed by traditional sequencing.
The data are sparser and less densely sampled than most of the artificial
examples analyzed above, so any estimates are necessarily rather imprecise.
Furthermore, we do not know exactly when infection occurred or CTL selection
started. The days given in the above papers are relative to the date of
identification of the patient as being HIV infected. It has been estimated that
in a chronically infected patient, there are a total of
around $4 \times 10^7$ infected cells \citep{haase_quantitative_1996}. 
Hence, the population size is $N\approx 10^7$  
but might be larger during peak viremia or smaller
due to bottleneck effects or the myriad of factors influencing patient-to-patient
variation in viral load. We determined posterior distributions for
population sizes ranging from $N=10^5$ to $N=10^8$. The mutation rate was set to
$10^{-5}$ per day \citep{Mansky:1995p38971}. This value is appropriate
if only one escape mutation per epitope is available. If escape can
happen in many different ways, a higher rate of about $\mu=10^{-4}$ per
day should be used and we repeated the estimation with $\mu=10^{-4}$
finding similar results, see below. Both of these scenarios are observed \citep{henn_whole_2012}.
Recombination in HIV occurs but is not
modeled here since its rate is low \citep{Neher:2010p32691,Batorsky:2011p40107},
and it is expected to be less relevant for the strong escapes in large
populations. In large populations, recurrent mutation is often more effective
at accumulating escape mutations than recombination between two rare variants.
Nevertheless, the neglect of recombination can lead to overestimation of escape
rates; see above. Lastly, we assume that infection occurred $\tau=20$ days before
the patient was identified and the viral population sampled \cite{Goonetilleke:2009p42296}.

For each patient, we initially considered all nonsynonymous mutations that are eventually
sampled at high frequency as potential candidates for sequential escape mutants. 
Nearby mutations in the same epitope were combined into one escape. We refined
this list of candidates by considering only time points early in infection that were sampled with
more than 5 genomes per time point and only the first 3-6 earliest
strong escapes. All samples used had between 7 and 15 sequences.
The frequencies of these escape mutations and their linkage into
multi-locus genotypes in the 5' and 3' half of the genome, which were
sequenced independently, can be easily
determined from the alignment provided in
Salazar-Gonzales et al \cite{SalazarGonzalez:2009p35091}. Linkage information between
the 5' and 3' half genomes is missing but can in all cases be imputed
using the assumption of sequential escapes.  We ignored mutations whose frequency
does not increase monotonically such as pol80 in subject CH40.
Later in infection, there is extensive non-synonymous diversity and it is not 
feasible to fit a time course for most of these mutations.

In CH40 we considered samples at time points $t = 0$, 16, 45, 111, and 181 days and 
identified escape in six epitopes; the first escape occurs in nef185,
followed by three indistinguishable escapes at gag113, gag389, and vpr74
and two additional escapes in vif161 and env145. Following 
Ganusov et al \cite{ganusov_fitness_2011} the number in the epitope name refers
to the beginning of the 18-mer peptide covering the epitope. The mutation
at env145 was not analyzed in Ganusov et al \cite{ganusov_fitness_2011} and 145
is simply the number of the mutated amino acid in gp120.
The indistinguishable escapes gag113, gag389, and vpr74 are treated as
described in section 2.4.3 on unobserved intermediates (all three escapes are
assumed to have identical escape rates and only their seed times are
varied). Note that the fifth escape at epitope vif161 shows almost the
same escape pattern as the three indistinguishable escapes preceeding
it. The escape rates of gag113, gag389, vpr74 and vif161 should
therefore be interpreted with care.
In CH58 we considered samples at time points $t = 0$, 9, 45, and 85 days and identified
four escapes; the first escape is at env581 and the second at env830,
followed by nef105 and gag236. 
In CH77 we considered samples at time points 
$t = 0$, 14, and 32 days and identified four escapes, namely the first escape
in tat55 and subsequent escapes in env350, nef17 and nef73.

\begin{figure*}[htp]
\begin{center}
  \includegraphics[width=0.89\columnwidth]{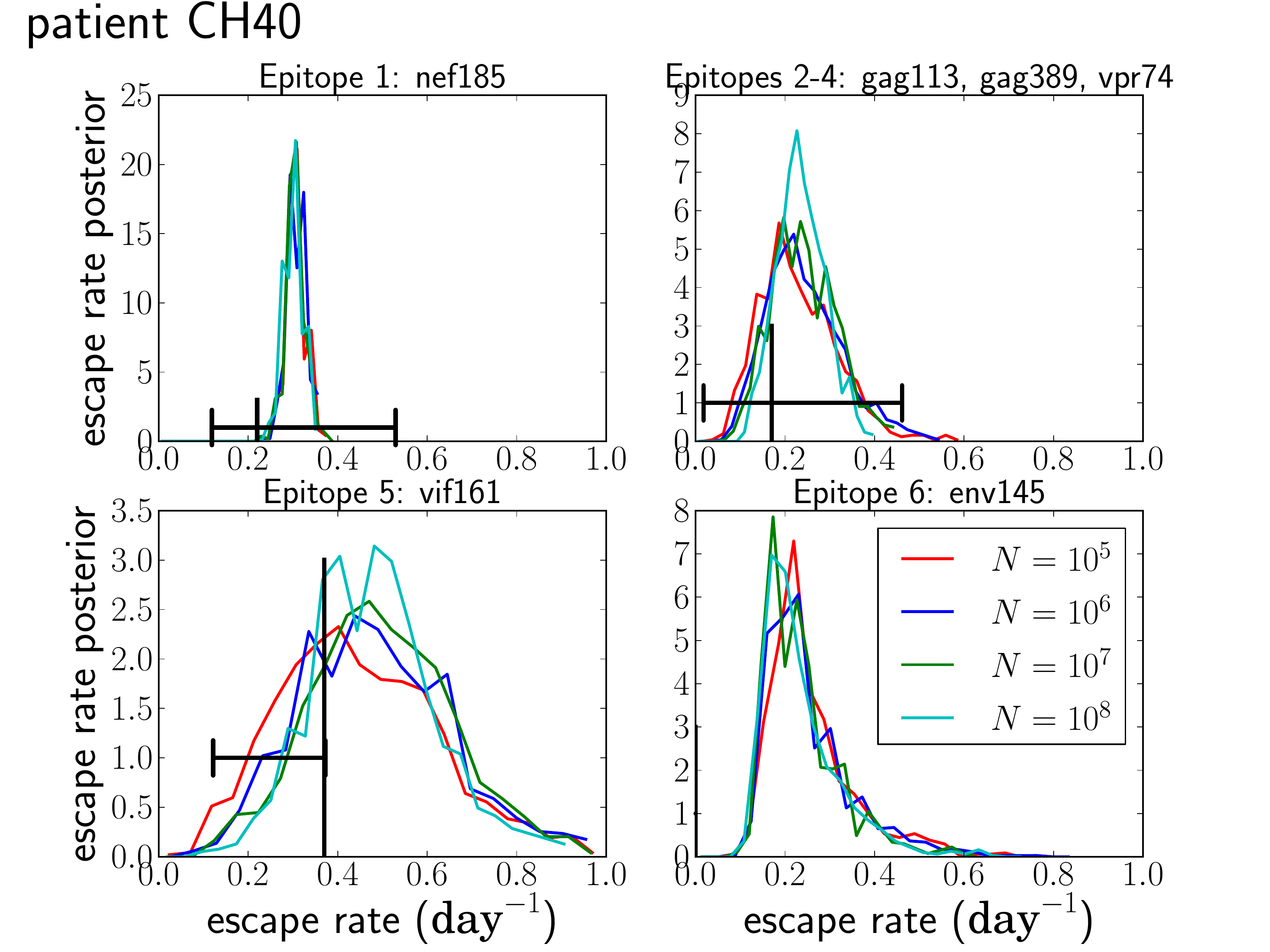}
  \includegraphics[width=0.89\columnwidth]{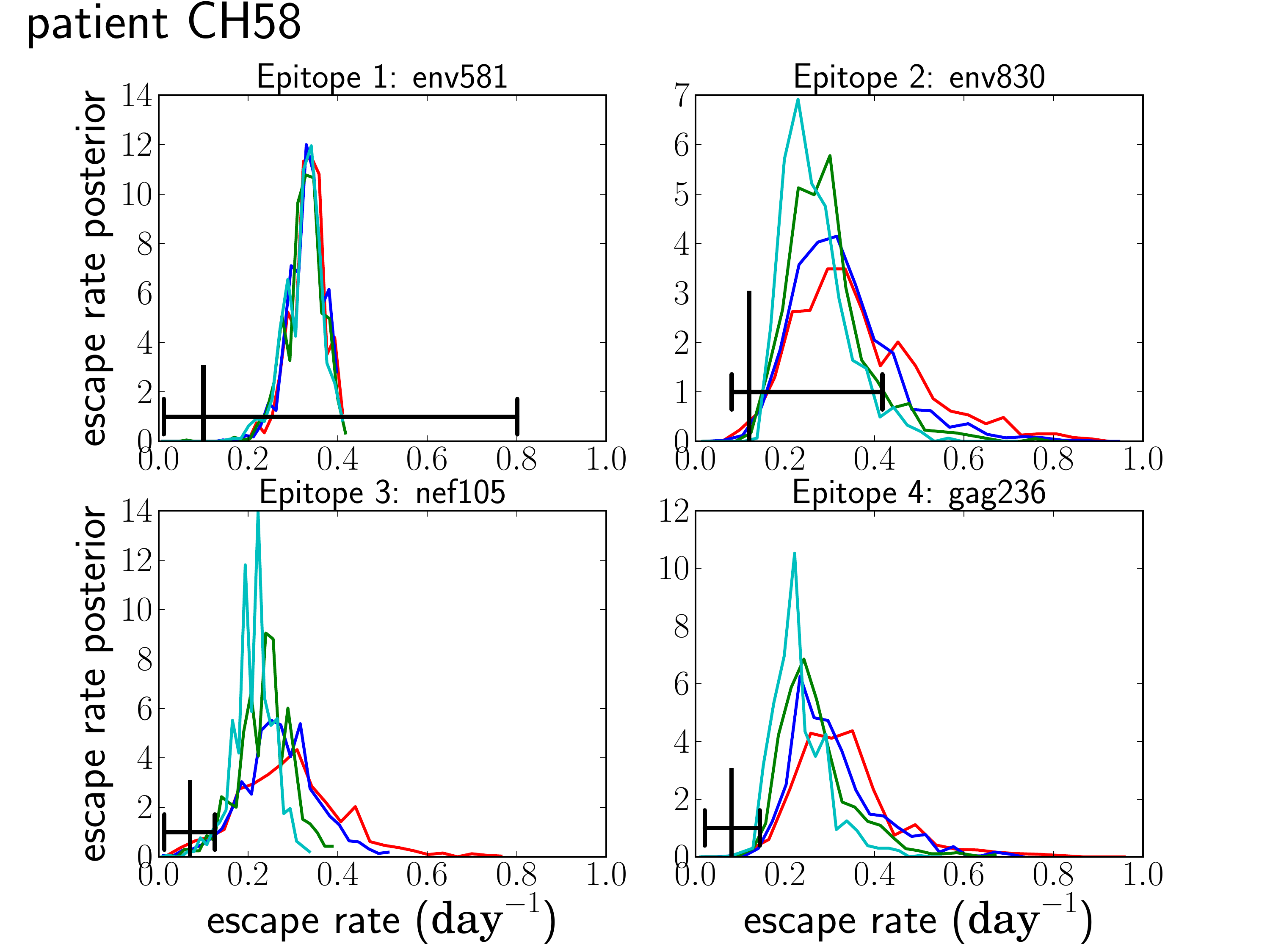}
  \includegraphics[width=0.89\columnwidth]{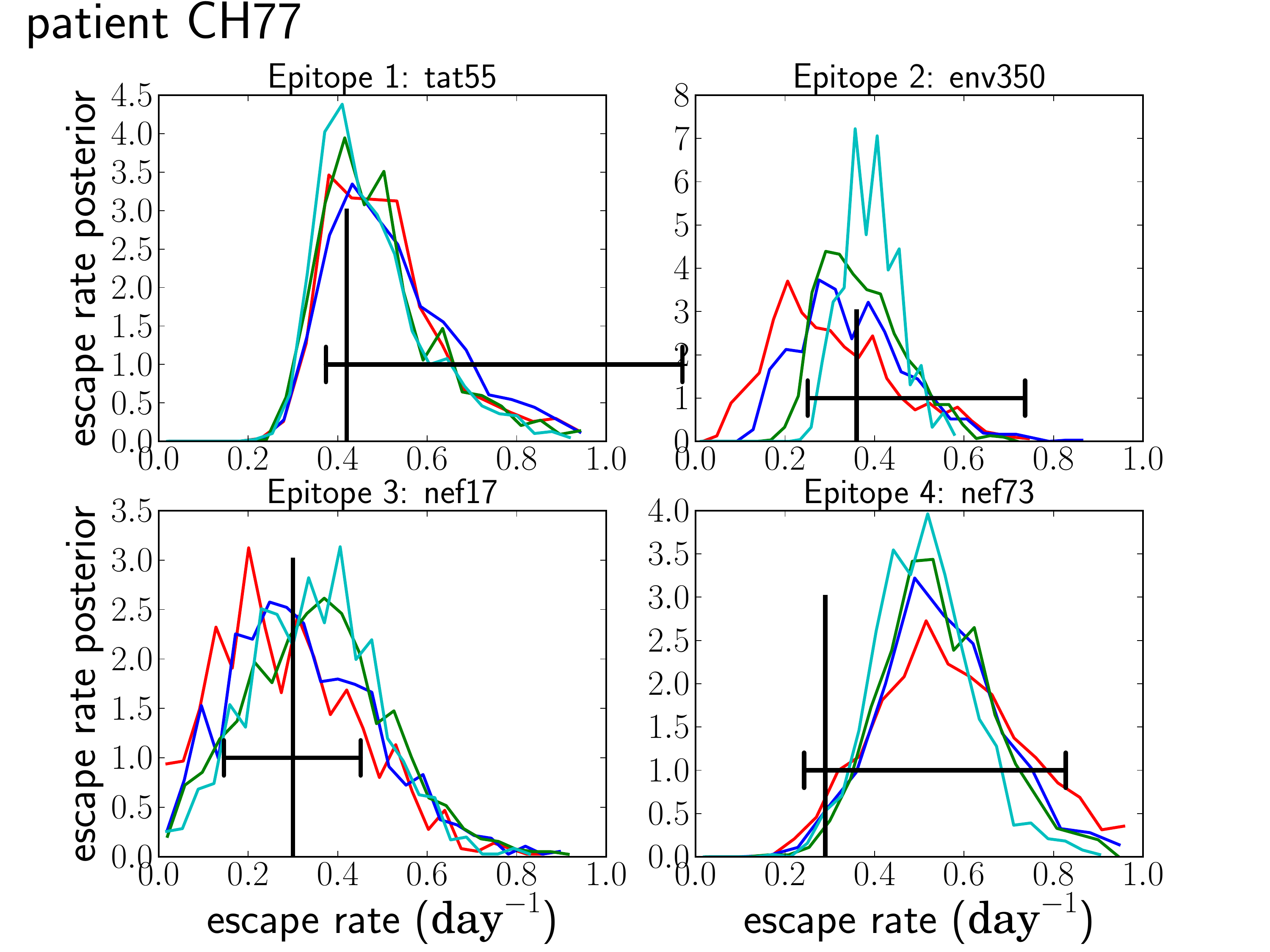}
  \includegraphics[width=0.89\columnwidth]{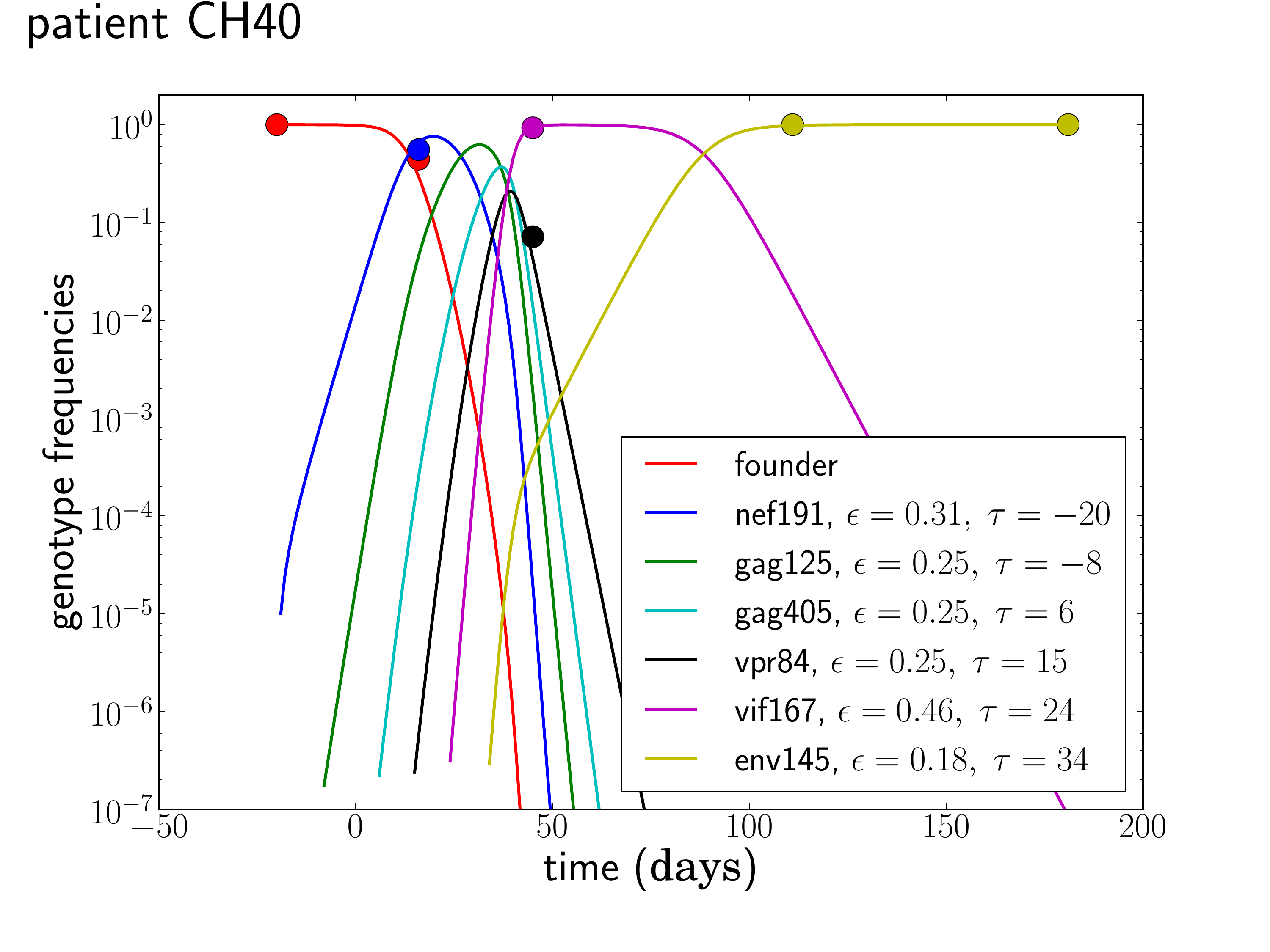}
  \caption[labelInTOC]{The posterior distribution of the escape rates for
  different population sizes. It is assumed that CTL selection starts 20 days
  prior to the date of identification, and the fitness prior has weight
  $\fitprior=10$. The black vertical and horizontal lines indicate the
  estimates and confidence intervals obtained in
  Ganusov et al \cite{ganusov_fitness_2011}. Note that the mutation env145 in
  CH40 was not analyzed in Ganusov et al \cite{ganusov_fitness_2011}. 
The lower right panel shows the most likely genotype trajectories for patient CH40 with 
parameters $N=10^{7}$ and $\mu=10^{-5}$. Each curve is labeled by an epitope, but should
be understood as the frequency of the genotype that has escaped at this and all previous epitopes. Note
that no data is available to differentiate epitopes gag113, gag389 and vpr74. For those, we assume an
arbitrary order and equal escape rates as explained in Sec.~2.4.3. }
  \label{fig:patients}
\end{center}
\end{figure*}

Given the above assumptions, we obtained estimates for the seed time and escape rate 
of each mutation. For each patient, we obtained initial estimates using a 
na\"ive single epitope fit for each mutation; then, we iterated
our multi-epitope fitting model five times. Next, we obtained posterior
distributions for the escape rates, all shown in \FIG{patients}, by performing a Markov chain Monte
Carlo simulation using the likelihood function given in \EQ{LH}. After obtaining our 
estimates, we randomly changed the escape rates in
increments of $\pm 0.01$ and the seed times by $\pm 1$, reevaluated the
likelihood, and accepted the change with probability $\exp(\Delta)$, where
$\Delta$ is the change in likelihood. The resulting Markov chain was run for
$10^6$ steps with samples taken every $1000$ steps.

Figure \ref{fig:patients} shows the posterior distributions of the escapes rate for
different epitopes in the three patients evaluated assuming a mutation
rate $\mu=10^{-5}$ per day. Larger
population sizes result in smaller estimates of the escape rates, as expected 
from \FIG{modeldeviation}A. 
The posterior distribution for the first escapes are often very tight,
but they depend on the time of the onset of CTL selection, which we have
set here to $T=20$ days prior to the first sample. If we assume that the
time of the onset of CTL selection coincided with the first sample (i.e., $T=0$), the
estimates of escape rates of the first epitope $\fcoeff_1$ are around
$0.9$, while later escapes are almost not sensitive to the choice of $T$.

\begin{figure*}[htp]
\begin{center}
  \includegraphics[width=0.89\columnwidth]{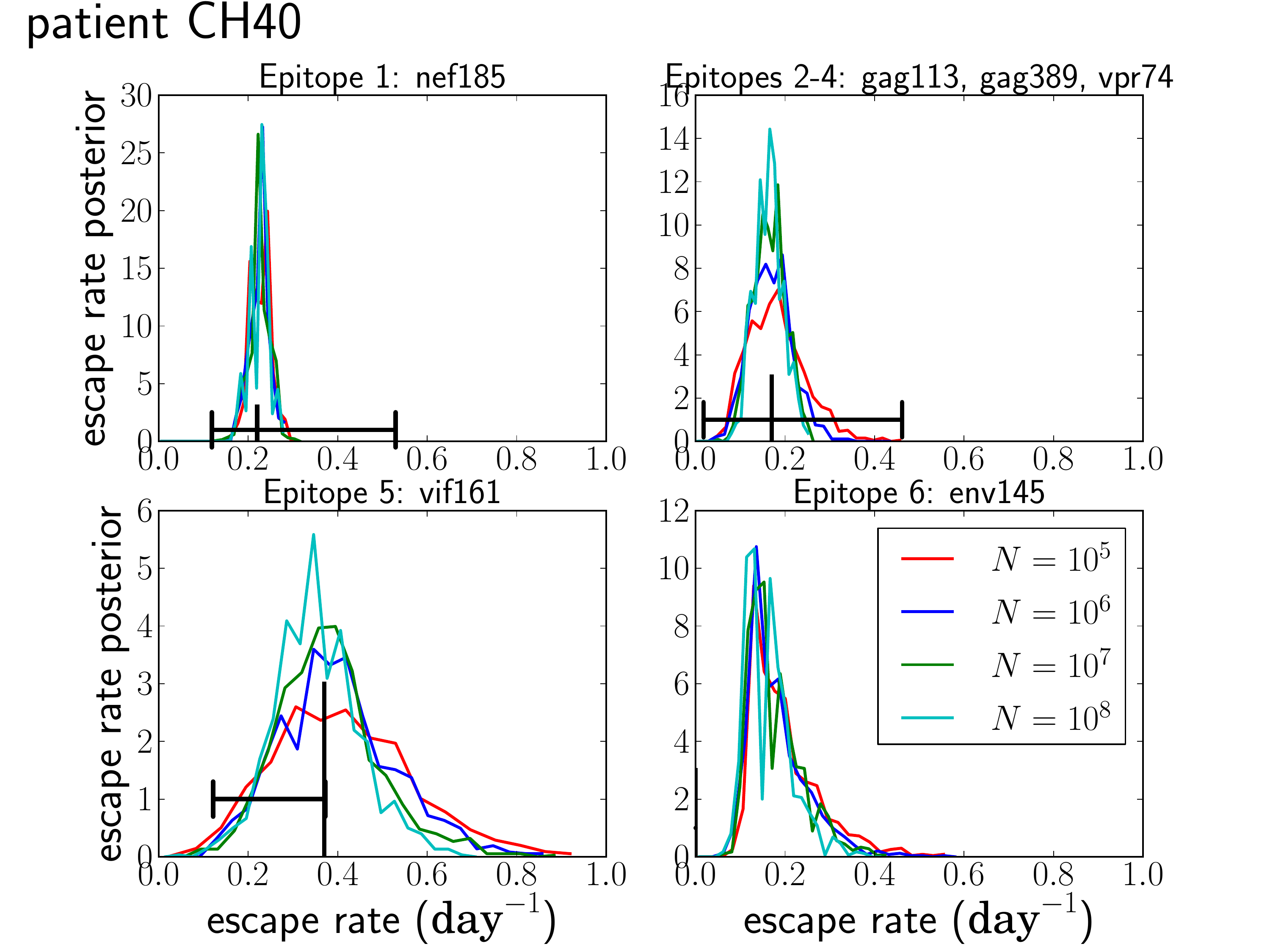}
  \includegraphics[width=0.89\columnwidth]{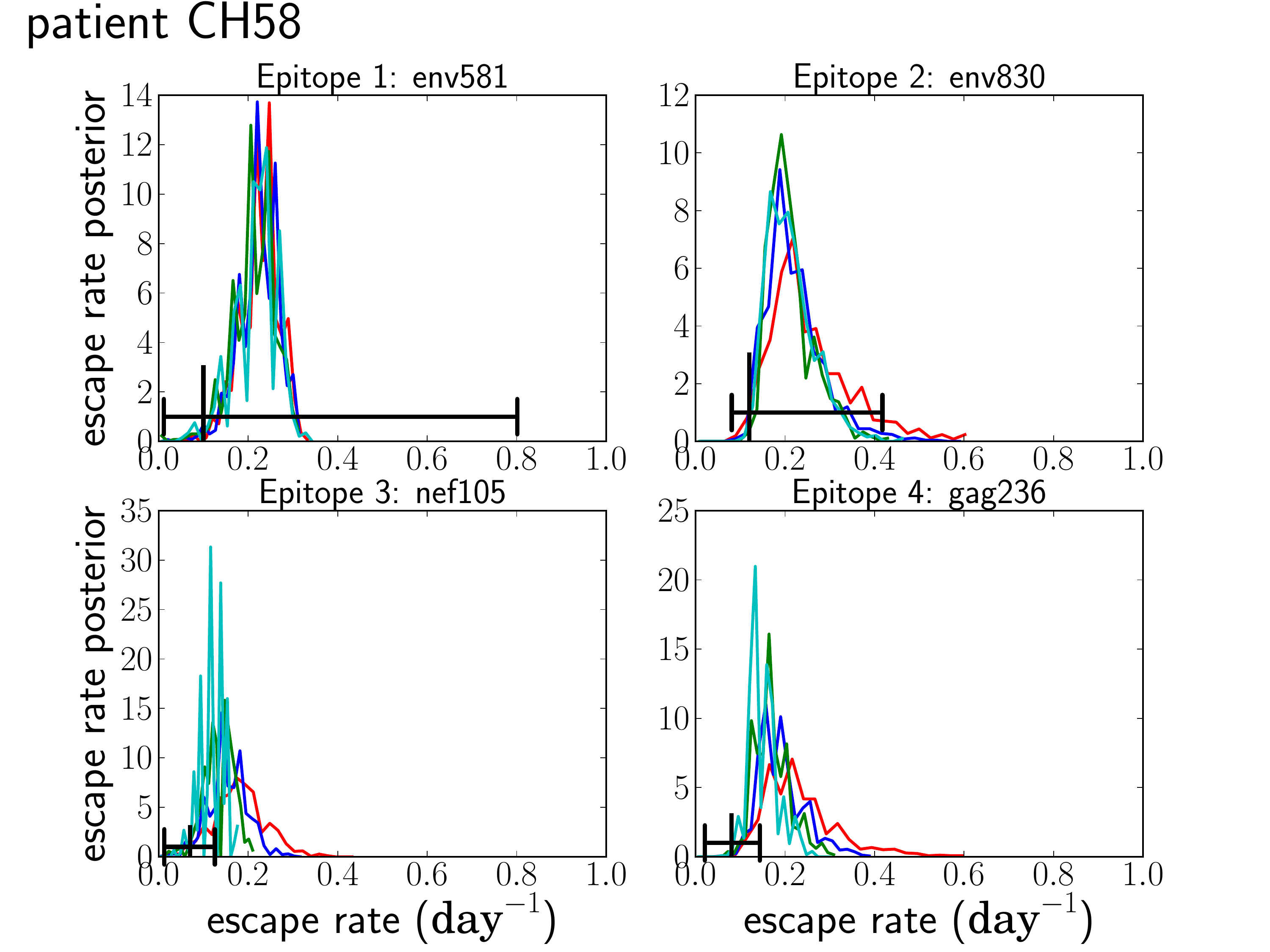}
  \includegraphics[width=0.89\columnwidth]{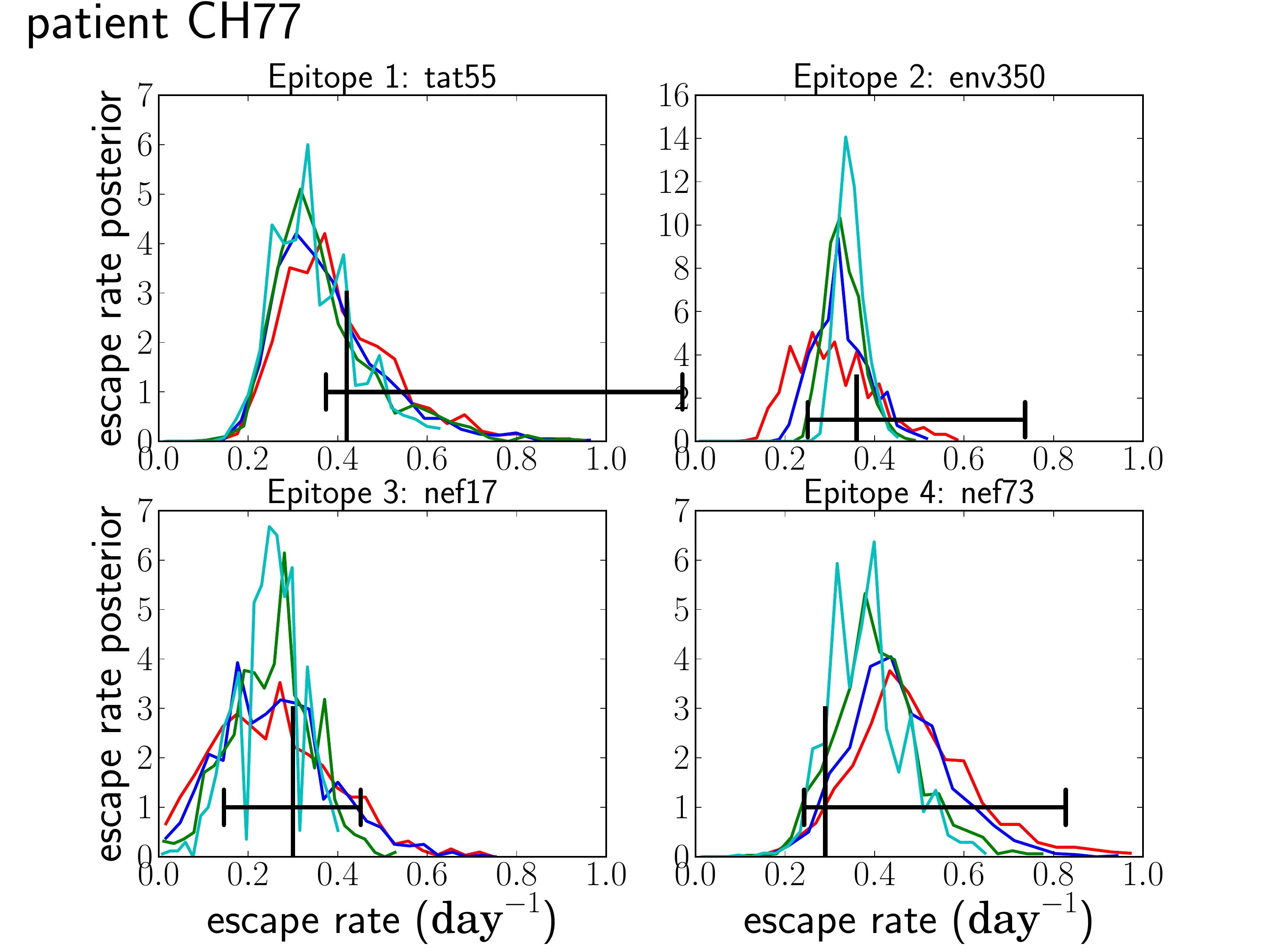}
  \includegraphics[width=0.89\columnwidth]{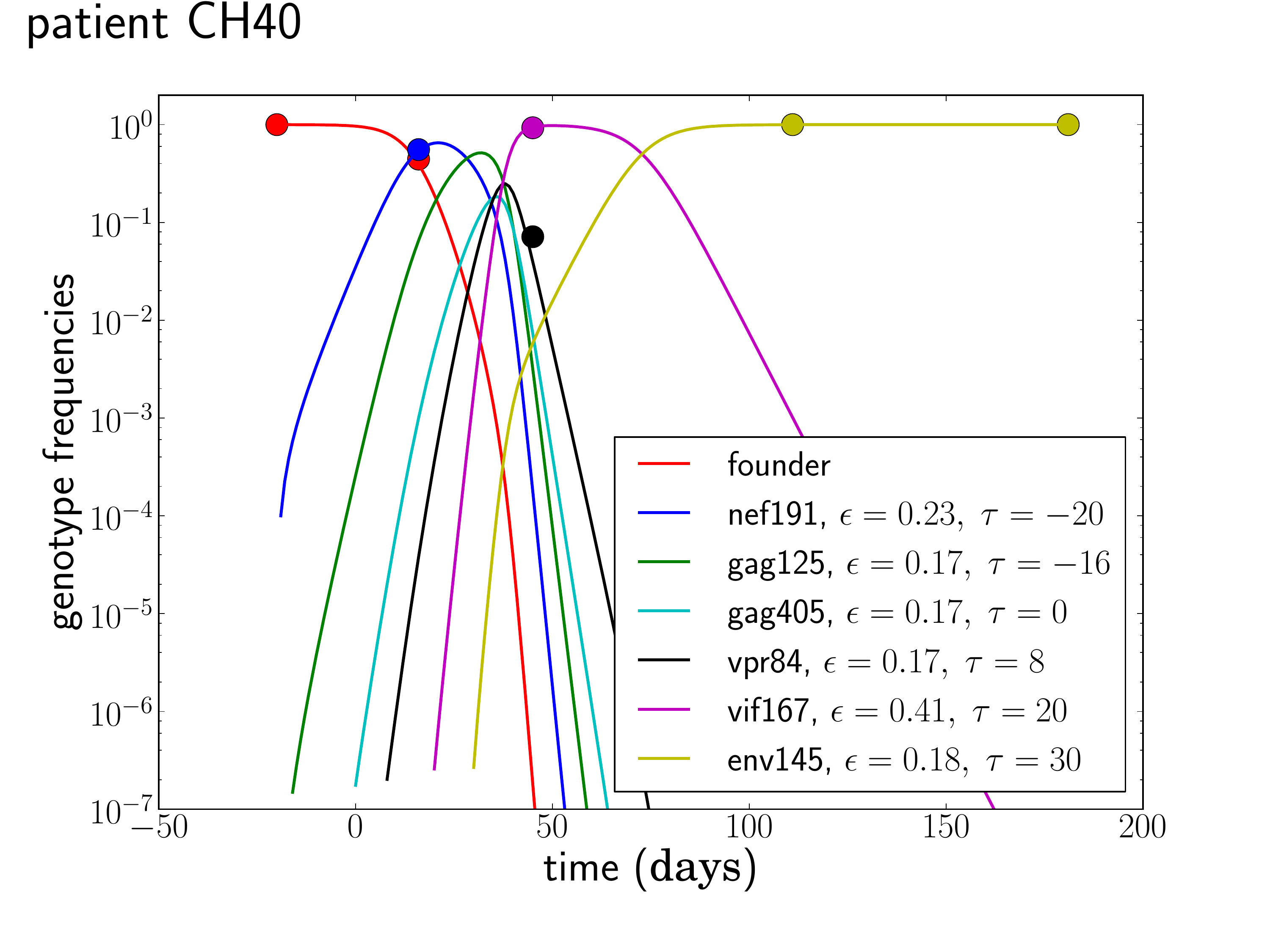}
  \caption[labelInTOC]{Posterior distributions of escape rate assuming a
    mutation rate $\mu=10^{-4}$ per day. See \FIG{patients} for other details. }
  \label{fig:patients_high_mut}
\end{center}
\end{figure*}

While the posterior distributions of escape rates of subsequent escape
rates are quite broad, they nevertheless suggest
that escape rates can be substantially higher than previously
estimated \citep{ganusov_fitness_2011,Asquith:2006p28003}.
Furthermore, the escape rate is not obviously negatively correlated with the
time of emergence during acute infection with HIV-1, at least for the
earliest four to six escapes.
The underlying reason for this is that selection on a late escape is only active after the
successful multiple mutant has been produced. In previous single epitope
estimates, selection was allowed to act on the mutant frequency from the very
beginning, resulting in a reduced estimate of the escape rate. Figure
\ref{fig:patients} also shows the inferred trajectories for the most
likely parameter combination for patient CH40. One clearly sees the
rapid rise and fall of multiple genotypes between the second and third
time point. Given the large number of genotypes involved and the little
data available, the escape rates estimated for this case are rather
noisy. But this analysis clearly shows that strong selection is
necessary to bring four mutations to fixation in just a few weeks.
We repeated the analysis of the patient data assuming a mutation rate of
$\mu=10^{-4}$ and show the results in \FIG{patients_high_mut}. The
overall picture is similar to what we found for $\mu=10^{-5}$ per day,
but escape rates tend to be lower.

\section{Discussion}
We have suggested a way to infer viral escape rates from time series data sparsely sampled
from the evolutionary dynamics of asexual or rarely sexual populations such as
HIV. We exploit the sequential nature at which escape mutations accumulate, which
allows us to constrain the times at which new escape mutations arose. These
constraints regularize the inference to a large extent, but additional stability
is gained by prioritizing small escape rates through an
exponential prior.

Rates of single escape mutations have so far been estimated by comparing the
time series data to a model that assumes logistic growth of the mutation with a
constant rate. This approach has been used to analyze the intra-patient dynamics
of recombinant HIV \citep{Liu:2002p20788}, drug resistance
\citep{Paredes:2009p40058,Bonhoeffer:2002p40059}, and CTL escape dynamics
\citep{ganusov_fitness_2011,Asquith:2006p28003,ganusov_estimating_2006,asquith_vivo_2007,petravic_estimating_2008}.
While these methods work well if
each mutation is sampled multiple times at intermediate frequencies, they
provide very conservative lower bounds when data is sparse.
Furthermore, they ignore the effects of competition between escapes at different
epitopes and assume that each epitope can be treated independently.
Since the recombination frequency in HIV is low
\citep{Neher:2010p32691,Batorsky:2011p40107,Josefsson:2011p42531}, this can be a
poor approximation. Our method improves on previous methods on both of these
counts. We explicitly model the competition between escape mutations. This
competition places constraints on the times at which genotypes with multiple
escapes first arise (double mutants arise only after the single mutants), which
makes the inference more robust and the lower bound tighter. 

A related method to estimate CTL escape rates has been proposed by
Leviyang \citep{leviyang_computational_2013}, who modeled multiple
escape mutations by an escape graph that is traversed by the viral
population. Combining these two approaches, intra-epitope competition
as modeled in \citep{leviyang_computational_2013} and the between
epitope competition studied here would be an interesting extension.
Similar ideas have been developed in the context of mutations in cancer
or evolution experiments \citep{illingworth_method_2012}.

While previous methods neglect interactions between epitopes altogether
-- equivalent to assuming very rapid recombination -- our method ignores
recombination during the inference. By comparison with simulations that
include recombination, we have shown that neglecting recombination can
result in overestimation of the escape rates by roughly 30\% at plausible
recombination rates of 1\%
\citep{Neher:2010p32691,Batorsky:2011p40107}. We also show that
neglecting recombination is less of a problem at higher mutation rates.
Note that neglecting recombination cannot explain the larger escape
estimates compared to previous studies. For patient CH58 we find escape
rates that are up to three-fold higher than earlier estimates 
\citep{ganusov_fitness_2011}, while we never see such a big deviation in
our sensitivity analysis. Furthermore, the errors made when neglecting
recombination for rapid early escapes are
comparable to the uncertainties that result from infrequent sampling
or more severe deviations of the model from reality, such as
time variable CTL activity. 

Reanalysis of CTL escape data from HIV using our method suggests that
CTL escapes are substantially more rapid than previously thought. Even
with a large prior against high escape rates ($\fitprior=10$), we
estimate that the escape rates of the first 4-6 escapes are on the order
of $0.3-0.4$ per day. The estimates at large population sizes are fairly
insensitive to the prior for population sizes of $10^{6}$ or
larger. Early in infection, it is plausible to assume that the relevant
size is $N=10^7$
\citep{coffin_hiv_1995,perelson_dynamics_1997,boltz_ultrasensitive_2012}.
If population sizes are small, relaxing the prior against high escape
rates results in larger estimates, which further supports our finding
that escape rates are often large and competition between escapes needs
to be modeled. Given the sparse data, we can only estimate parameters of simple models
and have to neglect many complicating features of HIV biology. Among
other factors, the rate at which escape mutations are selected depends
on the overall $R_0$ of the infection and CTL selection is probably time
variable \cite{ganusov_fitness_2011}. The estimated parameters therefore
represent time averaged effect escape rates.

The timing of escape has been shown to depend on epitope entropy
and immunodominance \citep{liu_vertical_2013}. However, we modeled only
the first four to six escapes in each patient from which rather little
information about differential timing can be obtained. In the case of
CH77, the first four escapes occurred within a month from the
identification of the patient. In patient CH58, it took roughly three
month for four escapes to spread and the estimated escape rates are
lower as expected. In the case of CH40 four of the six escapes show
almost or completely indistinguishable escape patterns and we have
little power to differentiate the escape rates at epitopes gag113,
gag389, vpr74 and vif161. Hence any meaningful correlation with
immunological features and epitope sequence conservation, i.e.,
low entropy, requires more data.

The proposed method to analyze multi-locus time series of adaptive
evolution could be useful in many context where the genotypic
compositions of large populations of viruses or cells can be
monitored over time. Whenever mutations occur rapidly enough that they
compete which each other, this competition has to be accounted for in the
analysis. Outside of virus evolution, possible applications include
the development of cancer and microbial evolution experiments.

\section{Materials \& Methods}

\subsection{Data Preparation.} 
Our fitting method uses counts $k_{ij}$ of genotypes $\gt_j$ at time
points $t_i$ to infer escape rates of individual mutations. The
procedure used to obtain successive genotype counts from sequence data
sampled from patients is outlined in the text. As input data, our
analysis scripts expect a white-space delimited text file with a format
shown in Table \ref{tab:data_example}. In addition, a separate file with
the total number of sequences at each time point can be provided. This
file is expected to have the same format as the matrix with the genotype
counts; see Table \ref{tab:data_example}. In absence of such a file, the
sample sizes at each time point are obtained by summing the genotype
counts.

\begin{table}
  \centering
  \begin{tabular}{|c|c|c|c|c|c|}
\hline
 time [days] & founder & 	env581 &	env830&	nef105&	gag236 \\ \hline
9	&5	&2	&0	&0	&0\\
45	&0	&0	&5	&3	&0\\
85	&0	&0	&0	&0	&8\\ \hline
  \end{tabular}
  \caption{Format of input data: The escape mutations are ordered
    first by the time of first observation and then by abundance. Each entry in the table
    in a particular column reports the number of times a sequence is
    observed containing the escape of that column and {\bf all}
    previous escape mutations. }
\label{tab:data_example}
\end{table}

To test our method,
artificial data $k_{ij} = k(\gt_j,t_i)$ were obtained from simulated
trajectories (generated by FFPopSim) by binomial sampling (with size $n_{i}$) at specified
time points $t_i$. Trajectory generation and sampling are implemented in the file
\texttt{model\_fit/ctlutils.py} at \url{http://git.tuebingen.mpg.de/ctlfit}; see below.

\subsubsection{Sequence data}
The HIV sequences for patients CH40, CH58 and CH77 where downloaded from
\url{http://www.hiv.lanl.gov/content/sequence/HIV/USER_ALIGNMENTS/Salazar.html}
\citep{SalazarGonzalez:2009p35091}. 

\subsection{Inference.} 
The inference procedure consists of \emph{initial guessing}, \emph{sequential
addition of escapes}, \emph{multi-dimensional refinement}, and estimation of
\emph{posterior distributions}. The implementation can be found in
\texttt{src/ctl\_fit.py}, with the C code for the likelihood calculation in
\texttt{src/cfit.cpp}.

\paragraph*{Initial guesses. } We produce initial guesses by single epitope
modeling. The frequency of each escape mutation, $\afreq_j$, grows logistically
with the escape rate \citep{ganusov_mathematical_2013}. We expect that only the frequency of the 
first escape mutation is significantly affected by mutational input, since it 
receives input from the abundant founder sequence, while the later escapes only 
receive mutational input from the previously escape genotype, which is still rare 
when the novel escape arises. Hence we only model the mutational dynamics of the
first escape. In a single epitope model, the frequency of the founder variant 
is one minus the frequency of the escape variant. The frequency of the escape variant
increases by $\mu (1-\nu_1)$ per day due to mutations from the founder, and decreases
by $\mu\nu_1$ due to further mutations to additional escapes. Combined with 
the logistic growth, the dynamics of  $\afreq_1$ is described by
\begin{equation}
\dot \afreq_1(t) = \fcoeff_1\afreq_1(1-\afreq_1) +\mu [1-2\afreq_1] \ .
\end{equation}
with initial condition $\afreq_1(0)=0$.
Note the difference between the allele frequency $\afreq$, which refers to a
particular escape mutation, and $\gtfreq$, which corresponds to frequencies of particular
multi-epitope genotypes. The above ODE has the solution
\begin{equation}
\afreq_1(t) = \frac{1}{2\epsilon_1}[\fcoeff_1-2\mu+R\tanh (\frac{\alpha+t}{2}R)]
\end{equation}
where $R = \sqrt{\fcoeff_1^2 + 4\mu^2}$ and $\alpha =
\frac{4\mu-2\fcoeff_1}{4\mu^2+\fcoeff_1^2}$\citep{ganusov_mathematical_2013}. 
The escape rate $\fcoeff_1$ is
determined by maximizing the likelihood (\EQ{LH}) using \texttt{fmin} from
\texttt{scipy} \citep{Oliphant:2007p25672}. 

The seed time $\tau_j$ of subsequent escape mutants $\gt_j$ is determined by
maximizing the seed time prior $Q(\tau_j|\gtfreq_{j-1})$ defined in
\EQ{seedtimes} using the previously determined $\gtfreq_{j-1}$. 
The frequencies of mutations are assumed to follow a logistic trajectory since
the genotype from which they receive mutational input is itself still at low frequency:
\begin{equation}
\afreq_j(t) =\frac{e^{\fcoeff_j(t-\tau_j)}}{e^{\fcoeff_j(t-\tau_j)} +
N\fcoeff_j} \quad j>1 \ .
\end{equation}
Again, we maximize the posterior probability, \EQ{LH}, to obtain an initial estimate of
$\fcoeff_j$.

\paragraph*{Sequential addition of escapes.} 
Given the initial estimates for the first escape, we now add subsequent escapes
to the multi-epitope model, which is formulated in terms of genotype counts
$k_{ij}$ and frequencies $\gtfreq_j(t)$. Note that the interpretation of
genotype counts depends on how many epitopes are modeled. For example, if we
model epitopes $1,\ldots,j$ out of a total of $L$ epitopes, counts for genotype
$j$ are $k_{ij} = \sum_{l=j}^{L}k_{il}$, i.e., we ignore all later escapes. 

If the added escape is unique, i.e., no other escape mutation has the exact
same temporal pattern, we calculate the likelihood on a $21\times 31$ grid
of escape rates and seed times; comp.~\FIG{sequential_fitting}. The grid spans values between between 0 and twice the
initial estimate for both the seed time and the escape rate. 
The most likely combination of seed time and escape rate is
chosen, and the procedure is repeated with the next epitope.

If multiple epitopes exhibit the same temporal pattern, we add them all at once,
constrain their escape rates to be equal, and assume they emerged in the order
listed in the genotype matrix. Since we now have to optimize one joint escape
rate and multiple seed times, we do not map the likelihood surface
exhaustively but rather perform a greedy search. We examine next-neighbor moves with
steps $\delta \tau = \pm 1$ day and $\delta \fcoeff = \pm 0.02$ per day, moves
which change all seed times by $\delta \tau$, and 20 moves in which all seed times and
escape rates are changed by $\delta \tau$ and $\delta \fcoeff$ with random
sign; the step that maximizes the likelihood is accepted.
This is repeated until no favorable move is found and further repeated with 
$\delta \fcoeff = 0.01$ and $0.001$ per day.

\paragraph*{Refinement.}
We then iterate sequentially over every epitope and optimize its seed times and
escape rates as described above, but with all other epitopes part of the
multi-epitope model. This typically leads to rather small adjustments and
converges rapidly.

\paragraph*{Posterior distributions.}
To determine the posterior distribution of the escape rates, we attempt to
change all seed times and escape rates by $\delta \tau = \pm 1$ day and $\delta
\fcoeff = \pm 0.01$ per day with random sign. The move is accepted with
probability $\max(1,\exp(-\Delta))$, where $\Delta$ is the difference in
log-likelihood before and after the change. We sample this Markov chain every
$1000$ moves and thereby map the posterior distribution of seed times and escape
rates.

\subsection{Usage}
\paragraph*{Availability}
All source code and scripts are available at
\url{http://git.tuebingen.mpg.de/ctlfit}. 

\paragraph*{Building}
The part of our method that is implemented in C and the python
bindings can be built using \texttt{make} and the Makefile provided in
the \texttt{src} directory. Prerequisites for building are
\texttt{python2.7}, \texttt{scipy}, \texttt{numpy}, \texttt{swig}, and
a \texttt{gcc} compiler. 

\paragraph*{Fitting}
Given a text file with genotype counts specified as shown in table
\ref{tab:data_example}, fitting is performed by calling the script
\texttt{fit\_escapes.py} with python. Parameters can be set via
command line arguments:
\begin{equation}
  \label{eq:command}
  \mathtt{python\ fit\_escapes.py\ \mbox{-}\mbox{-}input\ datafile}
\end{equation}
where \texttt{--input} specifies the file with the genotype
counts. Other parameters can be modified in a similar manner. Running
the script with the option \texttt{--help} prints a list of all parameters. The estimated
escape rates and seed times as well as the sampled posterior
distribution will be saved in the directory
\texttt{fit\_escapes\_output}, unless otherwise specified.

\section*{Disclosure/Conflict-of-Interest Statement}
The authors declare that the research was conducted in the absence of
any commercial or financial relationships that could be construed as a
potential conflict of interest.

\section*{Acknowledgements}
We are grateful for stimulating discussions with F.~Zanini. This work
is supported by the ERC starting grant HIVEVO 260686 (RAN) and in part by the National
Science Foundation under Grant No.~NSF PHY11-25915. This work was
performed under the auspices of the US Department of Energy under contract DE-
AC52-06NA25396 and supported by NIH grant AI028433, the National Center
for Research Resources and the Office of Research Infrastructure Programs (ORIP)
through grant OD011095 (ASP).

\bibliography{ctl_fitting}
\end{document}